\documentclass[aps,prx,onecolumn,superscriptaddress,nofootinbib, 11pt, floatfix]{revtex4-2}
\pdfoutput=1
\usepackage[utf8]{inputenc}
\usepackage[english]{babel}
\usepackage[T1]{fontenc}
\usepackage{amsmath, amsfonts, amssymb, amsthm}
\usepackage{comment}
\usepackage{tcolorbox}
\usepackage{hyperref}
\usepackage{graphicx}
\usepackage[caption=false]{subfig}
\usepackage{braket}
\newcommand{\ie}{{i.e., }}
\newcommand{\eg}{{e.g., }}

\newcommand{\N}{{\mathcal{N}}}

\newcommand{\A}{{\mathcal{A}}}
\newcommand{\B}{{\mathcal{B}}}
\newcommand{\C}{{\mathcal{C}}}
\newcommand{\D}{{\mathcal{D}}}
\newcommand{\Z}{{\mathcal{Z}}}
\newcommand{\mL}{{\mathcal{L}}}
\newcommand{\M}{{\mathcal{M}}}
\newcommand{\bZ}{{\mathbb{Z}}}
\renewcommand{\Vec}{{\rm Vec}}
\newcommand{\Rep}{{\rm Rep}}
\newcommand{\U}{{\rm U(1)}}

\usepackage{tikz}
\usepackage{lipsum}
\usepackage[utf8]{inputenc}
\usepackage[english]{babel}
\usepackage[T1]{fontenc}
\usepackage{amsmath}
\usepackage{amsfonts}
\usepackage{tikz}
\usetikzlibrary{decorations.pathreplacing}
\usepackage{hyperref}

\usepackage{tikz}
\usetikzlibrary{decorations.markings}

\tikzset{
  edge/.style={line width=1.1pt, line cap=round},
  outarrow/.style={
    postaction={decorate},
    decoration={markings, mark=at position 0.62 with {\arrow{Stealth}}}
  }
}
\usepackage{lipsum}
\usepackage{amsthm}
\usepackage{amssymb}
\usetikzlibrary{arrows.meta} 
\usetikzlibrary{arrows.meta,decorations.markings}
\usetikzlibrary{arrows.meta, positioning}

\usetikzlibrary{arrows.meta,decorations.pathmorphing,decorations.markings,backgrounds,positioning,fit,petri}

\newtheorem{theorem}{Theorem}
\newtheorem{prop}{Proposition}

\newtheorem{definition}[theorem]{Definition}

\begin{document}

\title{Average Categorical Symmetries in One-Dimensional Disordered Systems}

\author{Yabo Li}
\affiliation{Center for Quantum Phenomena, Department of Physics,
New York University, 726 Broadway, New York, New York 10003, USA}
\author{Meng Cheng}
 \affiliation{Department of Physics, Yale University, New Haven, Connecticut 06511, USA}
 \affiliation{School of Natural Sciences, Institute for Advanced Study, Princeton, NJ}
\author{Ruochen Ma}
\affiliation{Kavli Institute for Theoretical Physics, University of California, Santa Barbara, CA 93106, USA}
\affiliation{Department of Physics, University of California, Santa Barbara, CA 93106, USA}

\begin{abstract}
We study one-dimensional disordered systems with average non-invertible symmetries, where quenched disorder may locally break part of the symmetry while preserving it upon disorder averaging. A canonical example is the random transverse-field Ising model, which at criticality exhibits an average Kramers–Wannier duality. We consider the general setting in which the full symmetry is described by a $G$-graded fusion category $\B$, whose identity component $\A$ remains exact, while the components with nontrivial 
$G$-grading are realized either exactly or only on average.
     We develop a topological holographic framework that encodes the symmetry data of the 1D system in a 2D topological order $\Z[\A]$ (the Drinfeld center of $\A$), enriched by an exact or, respectively, average $G$ symmetry. Within this framework, we obtain a complete classification of anomalies and average symmetry-protected topological (SPT) phases: when the components with nontrivial $G$-grading are realized only on average, the symmetry is anomaly-free if and only if $\Z[\A]$ admits a magnetic Lagrangian algebra that is invariant under the permutation action of $G$ on anyons. When an anomaly is present, we show that the ground state of a single disorder realization is long-range entangled with probability one in the thermodynamic limit, and is expected to exhibit power-law Griffiths singularities in the low-energy spectrum. Finally, we present an explicit, exactly solvable lattice model based on a symmetry-enriched string-net construction. It yields trivial ground state ensemble in the anomaly-free case, and exhibits exotic low-energy behavior in the presence of an average anomaly.
\end{abstract}

\maketitle
\tableofcontents

\section{Introduction}

Symmetry plays an important role in quantum field theory and condensed matter physics. Beyond organizing phases and phase transitions within the Landau symmetry-breaking paradigm, symmetries can be associated with quantized invariants that constrain long-distance physics non-perturbatively \cite{2023symmetryreview}. In particular, a symmetry with a 't~Hooft anomaly forbids any symmetric state from being short-range entangled (SRE), thereby precluding a unique symmetric gapped ground state of a local symmetric Hamiltonian. A closely related concept is that of symmetry-protected topological (SPT) phases, which can be understood as SRE states (equivalently, invertible topological field theories) in one higher dimension, whose boundaries host the anomalous symmetry action. The classification of SPT phases and 't~Hooft anomalies with various protecting symmetries has been one of the central achievements in quantum many-body physics over the past decade \cite{2013SPT,2015ARCMPSPT}.

In recent years, the concept of symmetry has been generalized in many directions; particularly relevant to this work are non-invertible symmetries \cite{2023TASIShao}. Unlike group-like symmetries, non-invertible symmetries are implemented by conserved operators without inverses and are naturally described by fusion categories. A simple example is the Kramers–Wannier (KW) duality in the 1D critical Ising chain. Non-invertible (categorical) symmetries can also exhibit anomalies that constrain the ground state to be nontrivial or protect SPT phases, analogous to group-like symmetries \cite{2019ThorngrenWang,2024clusterstate,Seiberg:2024gek}.

In realistic systems, where disorder is ubiquitous and may locally breaks the symmetry while preserving it only upon disorder averaging, it is natural to ask whether the topological features of symmetry, \eg SPT phases and anomalies, apply to such systems. This question is particularly relevant for categorical symmetries, since some of them intrinsically mix with lattice translation \cite{Seiberg:2024gek,2024sQCA}, which is typically preserved only on average in the presence of lattice impurities. Previously, the effects of average symmetry in disordered systems have been studied for group-like symmetries \cite{2022ASPT,2025ASPT,2025XuJian}; here we extend this program to 1D categorical symmetries.

To be precise, we consider a 1D lattice and an ensemble of local Hamiltonians $\{ H[h] \}$ with their respective pure ground states $\{ |\Psi[h]\rangle \}$. Here $h$ denotes the quenched disorder in the Hamiltonian, with probability distribution $P[h]$, which we assume to be spatially uncorrelated. A sample prepared in a single experiment can be viewed as one instance of this disorder ensemble. We consider the case in which the full categorical symmetry $\B$ has two parts: (i) the exact symmetry $\A$ (which forms a fusion subcategory) satisfied by each disorder realization $H[h]$ and (ii) the remainder, consisting of average symmetries that are explicitly broken in each realization by the disorder but under which the probability distribution $P[h]$ is invariant. We ask under which conditions, and in what sense, the ground-state ensemble $\{ |\Psi[h]\rangle \}$ is nontrivial.

In the main part of this work, we focus on the case where $\B$ is a $G$-graded fusion category (with $G$ a finite group), in which $\A$ is embedded as the identity component. We develop a topological holographic picture for this setting. As discussed in Sec.\ref{sec:holo}, the central idea is to model the 1D system of interest as a slab of a 2D topological order, hereafter referred to as the bulk. In particular, the bulk supports a symmetry-enriched topological (SET) order whose anyon content is described by $\Z[\A]$, \ie the Drinfeld center of $\A$, enriched by a global symmetry $G$. A crucial mathematical result \cite{2002ENO} states that there exists a unique $G$-SET such that, after an appropriate choice of top boundary condition, the gapped excitations on the top boundary—including confined anyons and boundary terminations of $G$ domain walls—form the fusion category $\mathcal{B}$. Braiding with these gapped objects implements the global $\B$ symmetry of the original 1D system. In the slab setup, all phases in the original 1D system can be realized by distinct choices of boundary conditions for the 2D $G$-SET on the bottom boundary\footnote{Familiar readers may recognize that our holographic framework is an ungauged version of a $\Z[\B]$ slab.}. An advantage of our framework is that the clean and disordered settings—in which the part of $\B$ with nontrivial $G$-grading is exact or average—admit a simple bulk correspondence: the topological order $\Z[\A]$ is enriched by an exact or, respectively, an average $G$ symmetry.

In this picture, whether the categorical symmetry $\B$ is anomalous is equivalent to whether one can choose a gapped bottom boundary for which $\B$ remains unbroken. In the clean setting, where the entire $\B$ is exact, we identify three obstructions to such a gapped bottom boundary—and hence three possible anomalies. By contrast, when the part of $\B$ with nontrivial $G$-grading is only an average symmetry, only a single obstruction remains to realizing a short-range–entangled disorder ensemble (Def.~\ref{def:SREensemble}); we refer to this as the \emph{average anomaly}: $\B$ is anomalous if $\Z[\A]$ does not admit a magnetic Lagrangian algebra of anyons—physically, an $\A$-unbroken gapped boundary condition of $\Z[\A]$ (equivalently, an $\A$-SPT from the 1D perspective) that is invariant under anyon permutations by $G$. This provides a highly computable obstruction to a fusion category being anomaly-free, applicable in both clean and disordered settings. When this obstruction vanishes, distinct $\B$ average SPT phases are in one-to-one correspondence with distinct $G$-invariant magnetic Lagrangian algebras of $\Z[\A]$.

Subsequently, we study the physical consequences of the average anomaly. In Sec.\ref{sec:anom-fluxinsertion}, using a modified flux-insertion argument, we show that when the full symmetry $\B$ is anomalous, the ground state of a single disorder realization is long-range entangled (LRE) with probability one in the thermodynamic limit. Intuitively, each disorder realization can be viewed as a static configuration of $G$ domain walls—determined by the quenched disorder pattern—which, given spatially uncorrelated disorder, proliferate throughout the system probabilistically. Our result implies that a single sample can be SRE with nonvanishing probability only if, upon crossing a $G$ domain wall, the system remains within the same $\A$-SPT phase. We expect that a system with a nontrivial average anomaly generally exhibits power-law Griffiths (rare region) singularities at low energies, with a paradigmatic example being the random transverse-field Ising chain \cite{FisherRTFI1} with an anomalous average KW symmetry, which flows to an infinite-randomness fixed point.

Finally, when $\Z[\A]$ admits a $G$-invariant magnetic Lagrangian algebra, we prove that the full symmetry $\B$ is indeed anomaly-free. We do so by explicitly constructing an ensemble of solvable disordered Hamiltonians that satisfy the respective exact and average symmetry, based on the symmetry enriched string-net model \cite{ChengPRB2017,HeinrichPRB2016}. We show that, in this ensemble, every realization has a unique gapped ground state, indicating that the symmetry is anomaly-free. A similar argument also implies that any categorical symmetry with no exact component is anomaly-free.
In contrast, applying the same construction to the anomalous case yields an ensemble with nontrivial features—a typical instance has extensive ground-state degeneracy, and the probability that the ground-state can be prepared by a circuit of any fixed finite depth is exponentially small in the thermodynamic limit.

The rest of the paper is organized as follows. In Sec.~\ref{sec:exaave}, we review exact and average symmetries and the consistency condition relating them. We also define a short-range entangled ensemble; when such an ensemble is forbidden by a symmetry, this signals an average anomaly. In Sec.~\ref{sec:KWexample}, we present a simple example of an anomalous average symmetry—the average Kramers–Wannier duality—and analyze its consequences for the ground-state ensemble. In Sec.~\ref{sec:holo}, we introduce the topological holography framework, which maps the 1D system with categorical symmetry to the boundary of a 2D (average) SET. In Sec.~\ref{sec:modulecat}, we construct solvable lattice models with the desired symmetries and prove that certain average categorical symmetries are anomaly-free. We also discuss how the same construction, when applied to anomalous symmetries, yields ensembles with nontrivial low-energy behavior.

\section{Exact and average symmetries}

\label{sec:exaave}

First we review the notion of exact and average ordinary symmetries \cite{2022ASPT}. Suppose we start from a quantum lattice model with Hamiltonian $H_0$, which has symmetry $\tilde{G}$. The corresponding unitary transformations are denoted as $U_{\tilde{g}}$ for $\tilde{g}\in \tilde{G}$. Being a symmetry, it means that $U_{\tilde{g}}H_0=H_0U_{\tilde{g}}$ for all $\tilde{g}\in \tilde{G}$. Now we turn on quenched disorder and study the ensemble:
\begin{equation}
    H[h] = H_0+\sum_i h_i^\alpha O_i^\alpha.
    \label{eq:hamiltonianensemble}
\end{equation}
Here $O_i^\alpha$ are local operators supported near site $i$ and $h_i^\alpha$ are random variables following a probability distribution $P[h]$. For each disorder realization, $H[h]$ can be written as a sum of local terms whose supports have diameter bounded above by a constant $R$. We also assume that (i) \(P[h]\) is homogeneous in space (e.g.\ translationally invariant), so that a thermodynamic limit exists; and (ii) $P[h]$ is short-range correlated, i.e., $\overline{(h_i^{\alpha})^* h_j^\beta}$ decays exponentially with $|i-j|$. Physical observables are calculated for each $h$ and then average over the ensemble.

Suppose $O_i^\alpha$ are invariant under a normal subgroup $A\subset \tilde{G}$. Because $A$ is normal, we have the quotient symmetry $G=\tilde{G}/A$. The operators $\{O_i^\alpha\}_\alpha$ transform nontrivially under $G$. 

Since for any given $h$, $A$ remains a symmetry of $H[h]$, we call $A$ the exact symmetry. However, under a $g\in G$ transformation, $H[h]$ is transformed into a different Hamiltonian $H[{}^g h]$. If $P[h]=P[{}^gh]$, then $G$ remains a symmetry for the whole ensemble, even though it is broken explicitly for each disorder realization. Such a $G$ is called an average symmetry of the ensemble. For example, a random field $h_i$ is odd under spin flip, $h_i \mapsto -h_i$, and spin flip becomes an average symmetry provided that $P(h)=P(-h)$. Average symmetries are ubiquitous in condensed matter systems. Practically, any spatial symmetry, such as lattice translation, is broken by random impurities but preserved as on average.

We now discuss how to generalize these notions to non-invertible symmetries. For simplicity, we only consider one-dimensional systems where the mathematical structure is relatively well-understood.


\tikzset{
    edge/.style={
      line width=1.2pt,
      postaction={decorate},
      decoration={markings, mark=at position 0.55 with {\arrow{Latex}}},
    },
    lab/.style={font=\normalsize}
  }

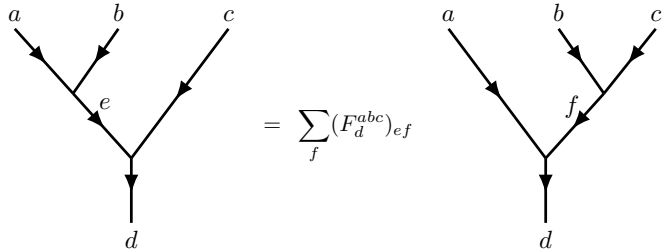
\begin{figure}[t]
\centering
\resizebox{0.54\columnwidth}{!}{%
\begin{tikzpicture}[baseline=(current bounding box.center)]

  \def\xL{-3.2}
  \coordinate (dL)  at (\xL,0);
  \coordinate (v1L) at (\xL,1.0);
  \coordinate (v2L) at (\xL-0.9,2.0);
  \coordinate (aL)  at (\xL-1.8,3.0);
  \coordinate (bL)  at (\xL-0.2,3.0);
  \coordinate (cL)  at (\xL+1.5,3.0);

  \draw[edge] (aL) -- (v2L);
  \draw[edge] (bL) -- (v2L);
  \draw[edge] (v2L) -- node[lab,above,pos=0.55,yshift=5pt] {$e$} (v1L);
  \draw[edge] (cL) -- (v1L);
  \draw[edge] (v1L) -- (dL);

  \node[lab, anchor=south] at (aL) {$a$};
  \node[lab, anchor=south] at (bL) {$b$};
  \node[lab, anchor=south] at (cL) {$c$};
  \node[lab, anchor=north] at (dL) {$d$};

  \node at (0,1.35) {${\Large \displaystyle=\ \sum_{f} (F^{abc}_{d})_{ef}}$};

  \def\xR{3.2}
  \coordinate (dR)  at (\xR,0);
  \coordinate (v1R) at (\xR,1.0);
  \coordinate (v2R) at (\xR+0.9,2.0);
  \coordinate (aR)  at (\xR-1.5,3.0);
  \coordinate (bR)  at (\xR+0.2,3.0);
  \coordinate (cR)  at (\xR+1.7,3.0);  

  \draw[edge] (aR) -- (v1R);
  \draw[edge] (bR) -- (v2R);
  \draw[edge] (cR) -- (v2R);
  \draw[edge] (v2R) -- node[lab,above,pos=0.55,yshift=1pt] {$f$} (v1R);
  \draw[edge] (v1R) -- (dR);

  \node[lab, anchor=south] at (aR) {$a$};
  \node[lab, anchor=south] at (bR) {$b$};
  \node[lab, anchor=south] at (cR) {$c$}; 
  \node[lab, anchor=north] at (dR) {$d$};
\end{tikzpicture}%
}
\caption{Illustration of the defintion of $F$-symbol}
\label{fig:fsymbol}
\end{figure}

\begin{figure}
\begin{center}
  \includegraphics[width=.70\textwidth]{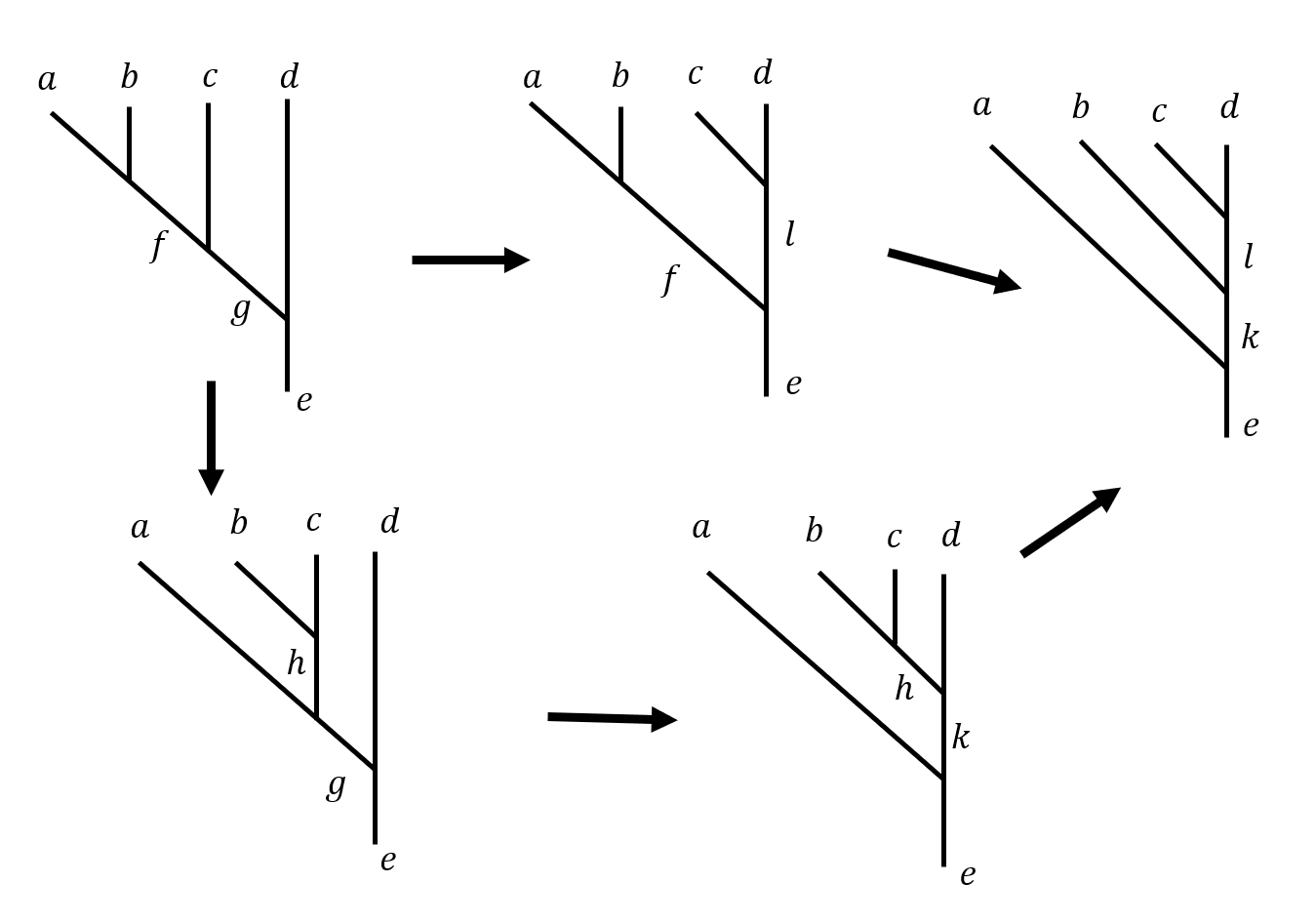} 
\end{center}
\caption{Illustration of the pentagon equation.}
\label{Fig:Fpentagon}
\end{figure}

Non-invertible symmetries are not described by groups, but rather by fusion categories. For the mathematical details, see Refs.~\cite{2006Kitaev,etingof2015tensor}. Here, we list only the defining data that are crucial for our discussion. A fusion category $\A$ is specified by a set of simple objects $\{ a \}$, and their fusion rules:
\begin{equation}
    a\times b = \sum_c N_{ab}^c c.
    \label{eq:fusion}
\end{equation}
The fusion coefficients $N_{ab}^c$ are non-negative integers which describe the number of distinct ways that $a$ and $b$ can fuse into $c$. The set of simple objects includes a unique unit object, denoted by $\mathbf{1}$, satisfying
\[
N_{c\,\mathbf{1}}^{\,a} = N_{\mathbf{1}\,c}^{\,a} = \delta_{a,c}
\quad\forall\,a,c\in\mathcal{A}.
\]
Moreover, for each simple object $a$ there exists a unique antiparticle $\bar{a}$ such that $\mathbf{1}$ appears in their fusion:
\[
N_{a\,\bar{a}}^{\,\mathbf{1}} = 1.
\]
The fusion of simple objects satisfies associativity up to a unitary\footnote{We focus on unitary fusion categories in this work.} $F$–matrix (Fig.~\ref{fig:fsymbol}). To ensure consistency, the $F$-symbols must obey the pentagon identity
\begin{equation}
(F^{fcd}_e)_{gl}(F^{abl}_e)_{fk}=\sum_h(F^{abc}_g)_{fh} (F^{ahd}_e)_{gk}(F^{bcd}_k)_{hl},
\label{eq:pentagon}
\end{equation}
which guarantees that the two fusion paths illustrated in Fig.~\ref{Fig:Fpentagon} yield the same result. For ordinary invertible symmetries, the $F$-symbols reduce to phases and this pentagon identity reduces to the 3-cocycle condition. Finally, each simple object has a real, positive quantum dimension defined by
\begin{equation}
d_a = \bigl|(F_a^{a\bar{a}a})_{\mathbf{1}\mathbf{1}}\bigr|^{-1},
\end{equation}
which, by Eq.~\eqref{eq:fusion}, satisfies
\begin{equation}
d_a\,d_b = \sum_c N_{ab}^{\,c}\,d_c.
\label{eq:dimfusion}
\end{equation}

Physically, a large class of examples of spin chains with fusion category symmetries are constructed using anyon chain models \cite{2007goldenchain} (see \cite{EvanJones} for a general discussion). We will discuss concrete models using generalized anyon chains in Sec.~\ref{sec:modulecat}. 

Denote by $\B$ the fusion category symmetry of the clean Hamiltonian $H_0$, and suppose that after turning on disorder, the remaining exact symmetry is another fusion category $\A$. Clearly, $\A$ needs to be a subcategory of $\B$. Naively, one might imagine that any symmetry transformation in $\mathcal{B}\setminus\mathcal{A}$ could serve as an average symmetry.  However, one must ensure that after applying such an average symmetry transformation to the Hamiltonian in Eq.\eqref{eq:hamiltonianensemble}, the resulting Hamiltonian—drawn from the same disorder ensemble—remains invariant under $\mathcal{A}$, i.e.\ that $\mathcal{A}$ continues to be an exact symmetry.  This condition is satisfied if the fusion rules of $\mathcal{B}$ are commutative, or, if $\mathcal{B}$ is an extension of the average symmetry by $\mathcal{A}$ \cite{2010extension}, which guarantees that $\mathcal{A}$ is “normal” in $\mathcal{B}$. Mathematically, we have

\begin{definition}
An average categorical symmetry is specified by a pair of fusion categories $\mathcal{A}\subset\mathcal{B}$, where $\mathcal{A}$ is the exact symmetry of each disorder realization, and the additional symmetries in $\mathcal{B}\setminus\mathcal{A}$ act only on average. These average symmetries satisfy the condition that, for every simple object $a\in\mathcal{A}$ and every simple object $x\in\mathcal{B}\setminus\mathcal{A}$, there exists $a'\in\mathcal{A}$ such that
\[
a\times x \;=\; x\times a'.
\]

\end{definition}

In this work, we will further restrict ourselves to a special class, which nevertheless covers many fusion category symmetries studied in literature so far. We assume that $\B$ is $G$-graded, where $G$ is a finite group, and $\A$ is embedded as the identity component. That is,
\begin{equation}
    \B=\bigoplus_{g\in G}\B_g,
\end{equation}
and $\A\cong \B_1$, where the subscript "1" denotes the identity element of $G$. More concretely, we denote the simple objects in $\B_g$ by $\{ a_g \}$. Then the fusion rules must be $G$-graded:
\begin{equation}
    a_g\times b_h = \sum_{c_{gh}\in \B_{gh}}N_{a_gb_h}^{c_{gh}} c_{gh}.
\end{equation}
We say that $\B$ is a $G$-graded extension of $\A$.

An important class of examples that have been widely studied in recent years is the Tambara-Yamagami (TY) category \cite{tambara1998tensor,WangYF2019}. It is a $\bZ_2$-graded extension of an Abelian group $\Vec_A$:
\begin{align}
    \B&=\B_1\oplus \B_g,  \\
    \B_1&=\Vec_A, \quad \B_g=\{\sigma\}.
\end{align}
With the fusion rules 
\begin{equation}
\begin{split}
    \sigma\times a &= a\times \sigma=\sigma,\\
    \sigma\times\sigma &=\bigoplus_{a\in A}a.
\end{split}
\end{equation}
The $F$–symbols in the Tambara–Yamagami category are fully classified.  For an Abelian group $A$, they are parametrized by a symmetric nondegenerate bicharacter $\chi\colon A\times A\to \U$ and a Frobenius–Schur indicator $\kappa=\pm1$, satisfying $(F_\sigma^{\sigma\bar{\sigma}\sigma})_{\mathbf{1}\mathbf{1}} = \frac{\kappa}{\sqrt{|A|}}$. 

More generally, for any $G$-graded fusion category $\mathcal{B}$ one can construct a new $G$-graded category $\mathcal{B}^{\omega}$ that shares the same simple objects and fusion rules as $\mathcal{B}$ but has modified $F$–symbols,
\begin{equation}
    \tilde{F}^{a_g,b_h,c_k}_{f_{ghk}} \;=\; \omega(g,h,k)\,F^{a_g,b_h,c_k}_{f_{ghk}},
\end{equation}
where $[\omega]\in H^{3}(G,\U)$ is a $3$-cocycle.\footnote{We work with the convention that $\omega$ is non-trivial only when all $g,h,k\neq 1$.} Physically, if $\B$ were an exact symmetry of the Hamiltonian, the associativity of fusion of local symmetry defects associated with $a_g,\,b_h,\,c_k$ would change by a phase $\omega$ \cite{2024LSMlattice}. In the disordered setting, however, the sectors $\mathcal{B}_g$ with $g\neq 1$ arise only on average: each individual realization is not symmetric and therefore does not support the corresponding local symmetry defects. As we will see later, these additional phase factors have no physical effect, and we regard $\mathcal{B}$ and $\mathcal{B}^{\omega}$ as representing the same average-symmetry category.

To formulate SPT order and anomalies in the disordered setting, we generalize the notion of an SRE state to an ensemble. This discussion largely parallels the treatment of average symmetry‐protected topological phases in the case of invertible symmetries \cite{2022ASPT,2025ASPT}. 
\begin{definition}
    Consider the family of ground states $\{|\Psi[h]\rangle\}$ of the Hamiltonians $\{H[h]\}$.  We say that this ensemble is short‐range entangled (SRE) if every $|\Psi[h]\rangle$ is symmetric under $\mathcal{A}$ and can be prepared by a finite‐depth circuit whose depth $\xi[h]$ is uniformly bounded across the ensemble.
    \label{def:SREensemble}
\end{definition}
We then define an average categorical symmetry to be \emph{anomalous} if it does not admit an SRE ensemble. In practice, we adopt a slightly relaxed version of Definition~\ref{def:SREensemble}: we call an ensemble SRE if there exists a finite \(\xi\) (independent of system size) such that the probability of states requiring a preparation circuit of depth greater than \(\xi\) vanishes in the thermodynamic limit.

Finally, if there exists an SRE ensemble respecting an average categorical symmetry, we say that the symmetry is anomaly-free.  Moreover, two SRE ensembles \(\{H_I,|\Psi_I\rangle\}\) and \(\{H'_I,|\Psi'_I\rangle\}\), with the same exact and average symmetry (where \(I\) labels each disorder realization and its ground state), are said to lie in the same phase if one can interpolate between them by a continuous deformation that (i) preserves the short‐range correlations of the disorder potential, (ii) respects the exact and average symmetries along the path, and (iii) keeps all ground states SRE throughout the deformation.  SPT phases protected by an average categorical symmetry are then defined as the equivalence classes of such SRE ensembles.

Before delving into the details of the classification, we present in the next section a simple example illustrating the physical consequences of an anomalous average categorical symmetry.

\section{An example: average KW duality}

\label{sec:KWexample}
A prototypical example of non-invertible symmetry is the KW duality in an Ising spin chain. Denote the Pauli X and Z operators on site $i$ by $X_i, Z_i$. The Ising symmetry is defined to be $X=\prod_j X_j$. Then the KW duality ${\mathsf  D}$ transforms the operators as 
\begin{align}
    X_j &\mapsto Z_jZ_{j+1}\\
    Z_jZ_{j+1} &\mapsto X_{j+1}.
\end{align}
Notice that $X_j$ and $Z_jZ_{j+1}$ generate all $\bZ_2$ symmetric operators, and ${\mathsf D}$ is a nontrivial locality-preserving automorphism of the algebra of symmetric operators \cite{2024sQCA}. On the other hand, ${\rm D}$  annihilates $\bZ_2$-odd operators, thus it is non-invertible. Within the $\bZ_2$-symmetric subspace, it follows from the transformation rules that ${\mathsf D}^2=T$, where $T$ is the unit translation. It is shown in \cite{Seiberg:2024gek} that the ${\mathsf D}$ is anomalous, that it forbids any symmetric SRE state. The ground state of a KW-dual Hamiltonian is either gapless (e.g. critical transverse-field Ising model) or spontaneously breaks the symmetry \cite{2018KWSSB}.

We now consider disordered systems where $\mathsf{D}$ becomes an average symmetry, while the Ising symmetry remains exact.
One concrete example is the random transverse-field Ising chain, given by the Hamiltonian:
\begin{equation}
    H = -\sum_i J_i Z_i Z_{i+1}-\sum_ih_i X_i,
\end{equation}
with both $\{J_i\}$ and $\{h_i\}$ random. When each coupling is independently drawn from the same distribution, the model has an exact $\mathbb{Z}_2$ Ising symmetry and an average KW duality obeyed at the level of the entire ensemble.

At the quantum critical point—identified by $\overline{\ln h} = \overline{\ln J}$, or by the presence of average KW duality symmetry—the random transverse–field Ising chain flows to Fisher’s \emph{infinite-randomness} fixed point \cite{FisherRTFI1,FisherRTFI2}.  Spins linked by the strongest bonds coalesce into correlated clusters whose typical size diverges under renormalization.  Integrating out modes down to an energy scale $\Omega$, we introduce the logarithmic RG variable $\Gamma = \ln(\Omega_0/\Omega)$, where $\Omega_0$ is a UV cutoff.  The remaining clusters span $\ell \sim \Gamma^{2}$ sites and carry magnetic moments $\mu \sim \Gamma^{\phi}$ with $\phi = (1+\sqrt5)/2$.  Correlations become ultra-broad: typical two-point functions decay as $\exp[-c\sqrt{r}]$, whereas the ensemble-averaged correlator follows the much slower power law
\[
\overline{\langle \sigma_i^{z}\sigma_j^{z} \rangle}
\;\sim\;
|i-j|^{-(2-\phi)},
\]
where $\phi = \frac{1+\sqrt{5}}{2}$. The dynamical exponent diverges, $z \to \infty$, as the lowest excitation gap closes as $\ln \Delta \sim -L^{1/2}$.

Although this long-distance behavior in the random transverse-field Ising model relies heavily on the specifics of the strong-disorder RG and cannot be deduced solely from kinematic constraints, one may still ask whether the existence of certain nontrivial behavior is guaranteed by the symmetry structure. A natural question is how likely a single disorder realization yields an SRE ground state. We argue below that such ``uninteresting'' ground states become exceedingly rare in the large-system limit. The key is to apply a partial symmetry transform (``flux insertion'') to the disordered Hamiltonian.

To begin, assume that for a given realization $H_I = \sum_X H_X$ in the ensemble (where $X$ denotes the support of each local Hamiltonian term), the ground state $|\Psi\rangle$ is short-range entangled, \ie it is $\bZ_2$ symmetric and can be prepared from a product state by a unitary circuit of finite depth $\xi$. In particular, for any $\mathbb{Z}_2$-symmetric operators $O_x$ and $O_y$, the connected correlation function obeys
$\langle O_x O_y \rangle_c:= \langle O_x O_y \rangle -\langle O_x \rangle\langle O_y \rangle\sim e^{-|x-y|/\xi}
$ in the state $|\Psi\rangle$ by the SRE assumption
, and likewise in $\mathsf D(|\Psi\rangle)$ because $\mathsf D$ is a locality-preserving automorphism of the $\bZ_2$-symmetric operator algebra \cite{2024sQCA}. Physically, $\mathsf{D}(|\Psi\rangle)$ lies in the $\mathbb{Z}_2$ spontaneous symmetry breaking (SSB) phase and is adiabatically connected to the GHZ state.

We now choose a large interval $A$ (with size $|A|$ much greater than $R$ and $\xi$), and consider another disorder realization $H_{I'} = \sum_X H_X'$, where
\begin{equation}
H_X' =
\begin{cases}
\mathsf D\bigl(H_X\bigr), & X \subset A,\\
H_X, & \text{otherwise}.
\end{cases}
\label{eq:fluxinsertion}
\end{equation}
That is, terms fully supported in $A$ are transformed by the average KW symmetry. Evidently, $H_{I'}$ is also a $\bZ_2$-symmetric local Hamiltonian. We denote the ground state of $H_{I'}$ by $|\Psi'\rangle$.  Since KW is an average symmetry, $H_{I'}$ belongs to the ensemble with the same probability as $H_I$.

Since the ground states of both $H_I$ and $\mathsf D(H_I)$ exhibit a finite correlation length $\xi$ with respect to the $\mathbb{Z}_2$-symmetric algebra—which here plays the role of a finite mass gap—$|\Psi'\rangle$ should have the same reduced density matrix as $\mathsf D(|\Psi\rangle)$ deep inside region $A$ and coincides with $|\Psi\rangle$ deep inside the complement $\bar{A}$.\footnote{Here we assume that, in a ground state with a finite correlation length, the local reduced density matrix is insensitive to symmetric modifications of the Hamiltonian far away. A rigorous proof of this assumption would require generalizing existing results on ``local perturbations perturb locally'' \cite{2022LPPL} to disordered systems, which may possess only a mobility gap rather than an energy gap \cite{2010hastingsdisorder}.
} This implies that $\mathbb{Z}_2$-odd operators develop correlations that do not decay with distance deep inside region $A$:
\begin{equation}
\langle \Psi'_I | Z_x Z_y | \Psi'_I \rangle 
\simeq \langle \Psi | \mathsf{D}\,Z_x Z_y\,\mathsf{D} | \Psi \rangle 
=O(1) \quad (x,y\in A) \,.
\label{eq:longrangesector}
\end{equation}
Physically, the partial KW transform in Eq.~\eqref{eq:fluxinsertion} generates a $\mathbb{Z}_2$-spontaneously broken sector within the $\mathbb{Z}_2$-disordered state $|\Psi_I\rangle$. Since region $A$ can be made arbitrarily large, the long-range correlation [Eq.\eqref{eq:longrangesector}] implies that the circuit depth required to prepare $|\Psi'\rangle$ can grow beyond any finite bound.

Finally, note that starting from an SRE state $|\Psi_I\rangle$ and for any finite length $l$, one can generate a state $|\Psi_I'\rangle$ with correlations extending beyond $l$ in infinitely many ways—e.g., by shifting the endpoints of region $A$ throughout the system or by creating multiple such regions, thereby producing multiple long-range–correlated domains. Importantly, all of these LRE states occur with the same probability as $|\Psi_I\rangle$ because $\mathsf{D}$ is an average symmetry of the ensemble: applying it to any large region does not change the realization probability. Therefore, in the thermodynamic limit, the probability of obtaining an SRE ground state $|\Psi_I\rangle$ vanishes.

From the above argument, we arrive at the following conclusion:
\begin{prop}
In 1D, a categorical symmetry with exact $\bZ_2$ and average KW duality is anomalous: in the thermodynamic limit, the $\mathbb{Z}_2$-symmetric ground state of a single realization in the ensemble is long-range entangled with probability going to 1.
\label{prop:averageKWanomaly}
\end{prop}

We conclude this section with several remarks:
\begin{enumerate}
    \item The average anomaly—which is \emph{absent} when considering only the invertible $\bZ_2$ symmetry—guarantees that in the large-system limit even a single disorder realization is long-range entangled. However, a deeper understanding of how average anomalies constrain physical observables (\eg disorder-averaged correlation functions) is still lacking. On physical grounds, we expect a 1D system with an average anomaly either (a) spontaneously breaks the exact symmetry; or (b) is critical, such that suitably chosen disorder-averaged or Edwards-Anderson correlation functions exhibit power-law behavior \cite{2026aanomaly}.

    \item As we will see later, all other anomalous average categorical symmetries yield the same physical consequence: in the thermodynamic limit, the ground state of any single disorder realization is long-range entangled.
    \item There is another way to understand Proposition \ref{prop:averageKWanomaly}. Since KW duality exchanges $\bZ_2$-ordered and disordered phases, a $\bZ_2$-symmetric SRE realization can be regarded as spontaneously breaking the average KW symmetry. By a refined Imry–Ma argument \cite{aizenman1990rounding}, in 1D such SSB of the average symmetry does not occur in any typical disorder realization, and its probability vanishes as the system size $L\to\infty$.
\end{enumerate}

\section{Anomaly via topological holography}

\label{sec:holo}

Topological holography gives a simple picture of a (1+1)d system that has fusion-category symmetry $\A$.  We model the system as a narrow strip (or ``sandwich") of a (2+1)d topological order. Mathematically, the (2+1)d bulk is described by a modular tensor category (MTC). In this case, the MTC is the Drinfeld center  of the fusion category $\A$, denoted by $\Z[\A]$.  The strip has two boundaries. On the top boundary, we impose the “canonical” boundary condition so that the category of confined defects (i.e., equivalence classes of gapped, localized excitations) is exactly $\A$. We refer to this as the “electric” boundary, which physically corresponds to the gapped boundary of $\Z[\A]$ where $\A$ is completely spontaneously broken. 

In this picture, local operators $\{O_\alpha^{(i)}\}$ correspond to bulk anyon lines $\alpha$ that condense on the top boundary and extend across the sandwich to the bottom boundary\footnote{The superscript $i$ labels different condensation channels when $\alpha$ is non-Abelian.}. These operators are organized by how they transform under the fusion-category symmetry: a boundary defect line $m\in\mathcal{A}$ acts on $O_\alpha$ through the \emph{half-braiding} between the endpoint of the bulk line $\alpha$ and the (top) boundary line $m$, which mixes the possible condensation channels according to half-braiding matrices. Physically, the symmetry action is realized by a boundary line that forms a ``lasso'' around the endpoint of $\alpha$ on the top boundary; the amplitude of this half-link is exactly the symmetry transformation on the local operator, $U_m\cdot O_\alpha^{(i)}=\sum_j \Psi^{m\alpha}_{ij}\,O_\alpha^{(j)}$. For a detailed discussion, see Ref. \cite{Huang:2023pyk}

In the holographic picture, gapped (1+1)d systems with $\A$ symmetry are realized by choosing different gapped boundary conditions of the (2+1)d theory $\Z[\A]$ on the bottom edge, sometimes called the dynamical boundary. It is shown that gapped boundaries of $\Z[\A]$ are in one-to-one correspondence with Lagrangian algebras \cite{2010lagrangian,Cong2017}.   
Mathematically, a Lagrangian algebra is a composite object in $\Z[\A]$ of the form
\begin{equation}
\mL \;=\; \bigoplus_{\alpha\in\Z[\A]} n_\alpha\,\alpha,
\end{equation}
where $n_\alpha\in\mathbb{Z}_{\ge0}$ and $n_\mathbf{1}=1$.  Physically, all anyon types $\alpha$ with $n_\alpha>0$ in $\mathcal{L}$ condense at the boundary, while all other anyons are confined. More precisely, objects in $\mathcal{L}$ have trivial topological spin and pairwise trivial mutual statistics, and their fusion closes within $\mathcal{L}$ and includes the vacuum. On the canonical reference top boundary, the condensed Lagrangian algebra is precisely the electric Lagrangian algebra, $\mL_e$, in $\Z[\A]$. For a detailed review, see Refs.~\cite{kong2014,Cong2017}.

Another important example for later discussion is the ``magnetic'' Lagrangian algebra $\mathcal{L}_m$, which is defined by the following property: the only anyon that appears both in $\mL_e$ and $\mL_m$  is the vacuum $\mathbf{1}$. Intuitively, it corresponds to condensing the ``flux-type'' excitations complementary to $\mathcal{L}_e$.  The gapped boundary associated with $\mathcal{L}_m$ preserves the full categorical symmetry. As a result, the categorical symmetry $\mathcal{A}$ is anomaly free if and only if there exists a magnetic Lagrangian algebra in $\mathcal{Z}[\mathcal{A}]$ \cite{2024zhangcordova}.

Now consider the case where the symmetry is $\B$, a $G$-graded extension of the fusion category $\A$ (with $G$ later treated as an average symmetry due to quenched disorder). A convenient topological holographic framework for phases with $\B$ symmetry is to model the (1+1)d system as a narrow slab whose (2+1)d bulk is $\Z[\A]$ enriched by an anomaly-free $G$ symmetry, described by a braided $G$-crossed extension of $\Z[\A]$~\cite{SET}, or a $G$-SET. The top boundary is chosen to be the canonical charge-condensing boundary of $\Z[\A]$. In this picture, the symmetry of the original 1D system corresponds to acting with $G$ in the lower half of the slab, thereby inserting a $G$-defect line (domain wall) parallel to the slab. In particular: 
\begin{itemize}
    \item Local operators fall into two classes: (a) operators supported on the bottom boundary, which transform in representations of $G$; and (b) anyon lines connecting the two boundaries, which transform according to their mutual braiding with the $G$-defect lines;
    \item For each $g\in G$, there can be topologically distinct types of $g$ defects. Because defect fusion follows the group law of $G$, the collection of all $G$-defect lines that act faithfully on local operators forms the $G$-graded fusion category $\B=\bigoplus_{g\in G}\B_g$. 
\end{itemize}
This construction provides a physical interpretation of the theorem in Ref.~\cite{ENO2009}, which states that there exists a unique $G$-SET such that $G$ domain walls form the $G$-graded fusion category $\B$. Equivalently, specifying $\B$ is tantamount to specifying how $G$ enriches $\Z[\A]$: it determines (i) anyon permutations (symmetry actions by braided autoequivalences), (ii) symmetry-fractionalization data on anyons, and (iii) possible stacking with $(2{+}1)$D $G$-SPT phases. Different choices of these data define distinct braided $G$-crossed extensions and hence distinct $G$-graded categories $\B$~\cite{SET}. Familiar readers may recognize that our holographic construction can be obtained from the standard holographic picture—where the 2D bulk is $\Z[\B]$, \ie the $G$ symmetry is gauged—by subsequently ungauging $G$. The advantage of our framework is that, as shown below, it is more convenient for disordered settings in which the $G$-graded components of $\B$ are symmetries only on average.

As a consequence, the question of whether $\B$ is anomalous reduces, in the slab picture, to whether one can choose a gapped bottom boundary on which $\B$ remains unbroken. More precisely, one must condense a set of anyons in $\Z[\A]$ at the bottom boundary such that (i) no anyon can tunnel from the top boundary and annihilate at the bottom (otherwise it would serve as a local order parameter for the breaking of $\B$); and (ii) the condensation at the bottom boundary does not break $G$. (i) simply requires that the condensation on the bottom boundary be a magnetic Lagrangian algebra in $\Z[\A]$. It is instructive first to ask whether a clean system with exact $\mathcal B$ symmetry is anomalous; the disordered case then simply follows.

\subsection{Clean setting}

We now ask whether, in a clean system with exact $\B$ symmetry, the symmetry is anomalous. Equivalently, can we find an $\A$-unbroken gapped boundary of the anyon theory--a magnetic Lagrangian algebra of $\Z[\A]$--whose condensation on the boundary preserves the global $G$ symmetry? There are three possible obstructions to realizing such an $\mathcal{L}_m$ \cite{Cheng:2020rpl, Bischoff:2018juy}:
\begin{enumerate}
    \item As an object in the MTC, $\mathcal{L}_m$ may be transformed by the $G$ action to a different Lagrangian algebra.  In that case, condensing $\mathcal{L}_m$ would break the $G$ symmetry.  Therefore, the $\mathcal{B}$ symmetry is anomalous if no $G$-invariant $\mathcal{L}_m$ exists in $\Z[\mathcal{A}]$.

    \item If $\mathcal{L}_m$ as an object is invariant under $G$, then one needs to check whether its algebra structure is also invariant under $G$. Physically, this means one can consistently assign “$G$-charges” to the anyons in the Lagrangian algebra. An obstruction appears when the condensed anyons carry projective $G$ quantum numbers -- then the condensation must break the $G$ symmetry. 
    \item Lastly, whenever the two obstructions above vanish, one can compute an $H^3(G,\U)$ obstruction arising from the $F$-move of three $G$ defects. If this obstruction is a nontrivial element of $H^3(G,\U)$, then the $G$ symmetry must have a 't Hooft anomaly given by the same group cohomology class. 
\end{enumerate}

Let us demonstrate this in a simple example in the case of ordinary group symmetry. 

In the first example, we consider  the $\bZ_2$ toric code enriched by a $G=\bZ_2$ symmetry. In the (1+1)d context,
this would correspond to $\bZ_2$ exact symmetry (so $\D=\Vec_{\bZ_2}$), and $G=\bZ_2$. We choose the total symmetry to be $\bZ_4$ -- equivalently,  in the toric code SET, the $e$ anyon carries half charge under $G$. 

 The $\bZ_2$ toric code has two types of gapped boundary, corresponding to $\mL_e=1+e$ and $\mL_m=1+m$. In the $e$-condensed boundary $\mL_e$, the $G$ symmetry must be spontaneously broken, since $e$ carries a projective quantum number under $G$. That is, while as an object $\mL_e$ is invariant under $G$, the second obstruction does not vanish.
 
 On the other hand, for the magnetic boundary with $m$ condensation, it is clear that the first two obstructions vanish, and we can assign a $\bZ_2$ charge to the condensed $m$. One option is that the $m$ carries no $\bZ_2$ charge. Then all obstructions vanish and the result is just a trivial state. We can also let $m$ to condense with an odd $\bZ_2$ charge. However, in this case there is an nontrivial $H^3(\bZ_2, \U)$ obstruction. In other words, the $\bZ_2$ symmetry is in fact anomalous.

Next we consider the bulk theory for $\Rep(D_8)$, viewed as an example of the TY category for $\Vec_{\bZ_2\times \bZ_2}$. In the bulk, we have a $\bZ_2\times \bZ_2$ toric code enriched by the following $\bZ_2$ symmetry:
\begin{equation}
    \rho: e_1\leftrightarrow m_2, \quad e_2\leftrightarrow m_1.
\end{equation}
One can show that the $\bZ_2$ symmetry generated by $\rho$ has a unique fractionalization class.  In particular, both $e_1m_2$ and $e_2m_1$ have no projective quantum numbers.

We choose the canonical electric boundary to be $\mL_e=1+e_1+e_2+e_1e_2$. There are two magnetic boundaries: 
\begin{equation}
    \mL_m^{(1)}=1+m_1+m_2+m_1m_2, \quad \mL_m^{(2)}=1+m_1e_2+m_2e_1+e_1m_1e_2m_2.
\end{equation}
We find
\begin{equation}
    \rho(\mL_m^{(1)})=\mL_e, \quad \rho(\mL_m^{(2)})=\mL_m^{(2)}.
\end{equation}
The magnetic boundary ${\cal L}_m^{(2)}$ is free from obstruction 1. Then we can have four assignments of $\bZ_2^2$ charges on its condensed anyons. However, the one where both $e_1m_2$ and $m_1e_2$ are charged has a $H^3$ obstruction \cite{Bhardwaj:2024qrf}. As a result, there are 3 $\Rep(D_8)$ SPT phases in total.

\subsection{Average anomalies}

\label{sec:anom-fluxinsertion}

We now extend the above argument to the disordered setting, where the $G$ symmetry is only realized on average. Our strategy is to test whether the above obstructions remain meaningful in the disordered setting. We start with obstruction~1, which has two cases: For any magnetic Lagrangian algebra $\mathcal{L}_m$, the average symmetry $G$ either (a) transforms $\mathcal{L}_m$ to an algebra that overlaps with $\mathcal{L}_e$, i.e., $G$ acting on the phase defined by $\mathcal{L}_m$ yields a phase where $\mathcal{A}$ is spontaneously broken, or (b) transforms $\mathcal{L}_m$ to a different magnetic Lagrangian algebra $\mathcal{L}_m'$ in $\Z[\mathcal{A}]$, meaning $G$ permutes among distinct $\mathcal{A}$-SPT phases. We argue that in both cases the average categorical symmetry remains anomalous, i.e., it does not admit an SRE ensemble.

\begin{prop}
Whenever $\Z[\A]$ does not have a $G$-invariant magnetic Lagrangian algebra, the average categorical symmetry $\B$ is anomalous: in the thermodynamic limit, the ground state of any single disorder realization is long-range entangled with probability 1.
\label{prop:averageanomaly}
\end{prop}
The argument proceeds analogously to that of Prop.~\ref{prop:averageKWanomaly}. Consider a realization $H_I=\sum_X H_X$ in the ensemble whose ground state $|\Psi\rangle$ is $\mathcal{A}$-symmetric and SRE; this corresponds to the gapped boundary of $\Z[\mathcal{A}]$ where $\mathcal{L}_m$ condenses. Let $\mathsf{D}\in\mathcal{B}$ (with nontrivial $G$ grading) be such that $\mathcal{L}_m$ is not invariant under $\mathsf{D}$. Then both $|\Psi\rangle$ and $\mathsf{D}(|\Psi\rangle)$ exhibit a finite correlation length $\xi$ for $\mathcal{A}$-symmetric local operators: for $|\Psi\rangle$ by its SRE nature, and for $\mathsf{D}(|\Psi\rangle)$ because $\mathsf{D}$ is a locality-preserving automorphism of the $\mathcal{A}$-symmetric operator algebra \cite{2023Jones}.

We now select a large interval $A$ whose length satisfies $|A|\gg R,\xi$ and define the partially transformed Hamiltonian
\begin{equation}
H_X' =
\begin{cases}
\mathsf D\bigl(H_X\bigr), & X \subset A,\\
H_X, & \text{otherwise}.
\end{cases}
\label{eq:fluxinsertionD}
\end{equation}
$H_{I'}$ is again an $\A$-symmetric local Hamiltonian. We denote the ground state of $H_{I'}$ by $|\Psi'\rangle$.  Since $G$ is an average symmetry, $H_{I'}$ belongs to the ensemble with the same probability as $H_I$.

Since the ground states of both $H_I$ and $\mathsf{D}(H_I)$ have finite correlation length $\xi$ for the $\mathcal{A}$-symmetric algebra, $|\Psi'\rangle$ shares the same reduced density matrix as $\mathsf{D}(|\Psi\rangle)$ deep inside $A$, and should coincide with $|\Psi\rangle$ deep inside $\bar{A}$.  In case (a), as in Proposition \ref{prop:averageKWanomaly}, this forces $|\Psi'\rangle$ to exhibit long-range correlations within $A$ due to spontaneous symmetry breaking of $\mathcal{A}$.  In case (b), it implies that $A$ and $\bar{A}$ lie in different $\mathcal{A}$-SPT phases. Since $\mathcal{A}$ is exact, each boundary of $A$ must host a projective edge mode. More precisely, a junction between distinct fiber functors, or $\A$-SPT phases, $f$ and $g$, is a simple object $X\in \mathrm{Fun}_{\mathcal A}(f,g)$ carrying a finite zero–mode space $V_X$ with $\dim V_X=\mathrm{FPdim}(X)>1$ whenever $f\not\simeq g$ \cite{kitaev2012,ENO2009,etingof2015tensor,inamura20241+}. If the full state remains $\mathcal{A}$‐symmetric, those two modes at $\partial A$ must be entangled to fuse into the identity module functor, producing long‐range correlations across a distance $|A|$.

Finally, starting from an SRE state $|\Psi_I\rangle$, one can generate states $|\Psi_I'\rangle$ with entanglement extending beyond any fixed length in infinitely many ways—each occurring with the same probability as $|\Psi_I\rangle$ due to the average symmetry of the ensemble. Therefore, in the thermodynamic limit, the probability of obtaining an SRE ground state $|\Psi_I\rangle$ vanishes. \qed

As an example, consider $\mathcal{B}=\mathrm{TY}(\mathbb{Z}_2)$, where the exact symmetry $\mathcal{A}=\mathrm{Vec}_{\mathbb{Z}_2}$ is extended by $G=\mathbb{Z}_2$ implementing the $e$–$m$ duality (the symmetry of Prop.~\ref{prop:averageKWanomaly}), in the holographic picture. $\Z[\mathcal{A}]$ admits a single magnetic Lagrangian algebra, $\mathcal{L}_m=1\oplus m$, which is not invariant under $G$ (case~(a)).  Consequently, $\mathcal{B}$ remains anomalous even when $G$ is imposed only on average.

By contrast, we claim that obstructions 2 and 3 are no longer relevant in the disordered setting. Recall that they apply only when one seeks a state that is both $G$-symmetric and SRE: vanishing of obstruction 2 ensures that one can consistently define the $G$-charge of the $\mathcal{L}_m$ condensate, while vanishing of obstruction 3 ensures the condensate is an SRE state respecting $G$ symmetry. Since in each disordered realization there is no exact $G$ symmetry—the ground state of a single instance is not a $G$-eigenstate—these obstructions no longer lead to an anomaly. 

Our conclusion regarding average anomaly can also be understood from a bulk perspective. In the presence of disorder, the 2D bulk is an SET enriched by an average $G$ symmetry, and whether the categorical $\B$ symmetry is anomalous in 1D maps to whether such an average SET admits an $\A$-symmetric gapped boundary. As discussed in Ref.~\cite{2025ASPT}, the permutation action of $G$ on anyons completely classifies distinct average $G$-SET phases. Consequently, the average anomaly of $\B$ is fully specified by the anyon-permutation data, which determines whether obstruction~1 is present. In contrast, the symmetry fractionalization data and the SPT class -- responsible for obstructions~2 and~3 -- are no longer topological invariants for average symmetry. Consequently, they drop out in the disordered setting.

Finally, whenever the symmetry is anomaly-free, the system admits a symmetric SRE ensemble—an ``average SPT'' phase. As in the clean case, there is a one-to-one correspondence between $\mathcal{B}$ average SPT phases and $G$-invariant magnetic Lagrangian algebras of $\Z[\mathcal{A}]$.

Before closing this section, we comment that a nontrivial average anomaly provides a topological origin for strong rare-region effects in disordered systems. To see this, let us label the phases permuted by the average symmetry by $\{\alpha,\, \beta,\,\ldots \}$. When the disorder is spatially uncorrelated, the probability of finding an interval of length $r$ that is uniformly in phase $\alpha$ is exponentially small, and scales as $\mathrm{Prob}(r)\simeq e^{-c r}$. On the other hand, the average anomaly ensures that the gap of such rare regions also decays exponentially with $r$: for case (a) in the first paragraph of this section, it arises from the degeneracy of the $\mathcal{A}$-broken ground states on the interval, while for case (b) it originates from the zero modes at the boundaries of the interval. The compensation of the two exponentials then leads to power-law Griffiths singularities at low energies. As discussed in Ref.~\cite{2019Vojta,2013Vojta}, when the clean, non-disordered part of the Hamiltonian is at a continuous phase transition (which is natural in $1d$, as the clean part is symmetric under an anomalous exact symmetry $\mathcal{B}$), this compensation gives rise to a strong power-law Griffiths singularity, characterized by a nonuniversal Griffiths dynamical exponent $z'$ in the low-energy density of states, \ie $\rho(\epsilon)\simeq \epsilon^{1/z'-1}$. Depending on whether the Harris criterion is violated ($\nu<2$, where $\nu$ is the correlation-length critical exponent of the clean continuous transition) or fulfilled ($\nu>2$), the corresponding disordered system may flow to an infinite-randomness regime with $z'\to\infty$ -- a paradigmatic example being the random transverse-field Ising chain discussed in Sec.~\ref{sec:KWexample} -- or remain governed by the clean critical point, with disorder generating only subleading Griffiths singularities characterized by a finite $z'$.

\subsection{No anomaly from average symmetry alone}

In this subsection, we show that if the system has only an average categorical symmetry $\mathcal{C}$ (with no exact symmetry), then for almost every disorder realization in the ensemble there is no obstruction to realizing an SRE ground state, implying that the symmetry is anomaly free. We first provide a topological holography argument and then present an explicit lattice-model construction.

The argument begins by considering the clean $\mathcal{C}$‐symmetric system in the ``sandwich” picture: the bulk on a strip is described by the Drinfeld center $\Z[\mathcal{C}]$, and the reference boundary is the canonical one obtained by condensing the electric Lagrangian algebra $\mathcal{L}_e$ in $\Z[\mathcal{C}]$.  We then choose the dynamical boundary to condense the same algebra $\mathcal{L}_e$, so that the resulting $1D$ system exhibits exact $\mathcal{C}$ symmetry which is fully spontaneously broken. Because of the $\mathcal{C}$-SSB, the ground state exhibits long‐range entanglement with a $|\C|$-fold degeneracy, where $|\C|$ is the rank of $\mathcal C$.

We now introduce charged disorder that reduces the exact $\mathcal{C}$ symmetry to an average one. In the sandwich picture, a local operator is represented by a bulk anyon line $\alpha$ that condenses (``terminates'') on the reference boundary and extends across the strip. The $\mathcal{C}$ symmetry acts on these local operators via their half‐braiding. The Hamiltonian for a given disorder realization is then described by Eq.\eqref{eq:hamiltonianensemble}, where \(H_0\) describes the clean system with exact $\mathcal{C}$ symmetry that is spontaneously broken, and \(O_i\) denotes the anyon tunneling operator at site \(i\). When the operators $\{ O_i \}$ span all anyon types in $\mathcal{L}_{e}$, there is no exact symmetry and the entire $\mathcal{C}$ is realized only on average. This symmetry-breaking disorder lifts the $|\mathcal{C}|$-fold ground-state degeneracy and produces a trivial ground state. In fact, in $d\le2$, the Imry–Ma theorem \cite{aizenman1990rounding,imry1975random} ensures that an initially ordered phase is unstable to arbitrarily weak, spatially uniform random charged perturbations; thus the ground state lacks long-range order in a typical disorder realization. Hence, an average symmetry by itself is always anomaly-free.

To further support our conclusion, we construct a solvable lattice model. The construction is based on the string-net model \cite{Levin05a} (for a review, see Ref.~\cite{levin20}). In short, given a fusion category $\C$, the string-net model is a systematic scheme that generates an exactly solvable Hamiltonian in $2$D whose ground state realizes the topological order $\Z[\C]$. In particular, the model is defined on a $2$D honeycomb lattice, where each edge hosts a local Hilbert space with orthonormal basis states $\{|a\rangle\}$, and $\{ a \}$ labels the simple objects of $\mathcal{C}$, referred to as “string types.” For simplicity, we illustrate with the Fibonacci categorical symmetry, and the generalization to other cases is straightforward. The Fibonacci category has two simple objects, $\mathcal{C} = \{1,\tau\}$, with the fusion rule
\[
\tau \otimes \tau = 1 \oplus \tau.
\]
This symmetry, when exact, is known to be anomalous—i.e., it forbids a unique, gapped, symmetric ground state \cite{2007goldenchain,2024zhangcordova}. One way to see this is to note that $\Z[\C]$, i.e., the doubled Fibonacci topological order, has only one Lagrangian algebra, $\mathcal{L}_e = \mathbf{1}\oplus \tau\bar\tau$, and admits no magnetic Lagrangian algebra.

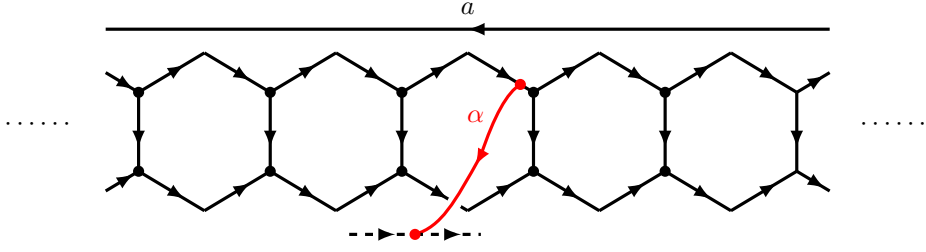
\begin{figure}[t!]
  \centering
  \begin{tikzpicture}[thick,scale=0.7, every node/.style={scale=1.0},
    arrow/.style={
    line width=1.2pt, 
    postaction={decorate},
    decoration={markings, mark=at position 0.7 with {\arrow{latex}}}
    },
    arrown/.style={
    line width=1.2pt, 
    postaction={decorate},
    decoration={markings, mark=at position 0.85 with {\arrow{latex}}}
    },
    arrowm/.style={
    line width=1.2pt, 
    postaction={decorate},
    decoration={markings, mark=at position 0.5 with {\arrow{latex}}}
    }
  ]
  \def\n{5}       
  \def\gap{2.5}     
  \def\h{1.5}     

  \foreach \i in {0,1,2,3,4} {
    \pgfmathsetmacro\x{\i*\gap}
    \fill[black] (\x,0) circle (3pt); 
    \draw[arrow] (\x,\h) -- (\x,0);  
    \fill[black] (\x,\h) circle (3pt);
  \draw[arrow] (\x,0) -- (\x+0.5*\gap,-0.5*\h);
  \draw[arrow] (\x+0.5*\gap,-0.5*\h) -- (\x+\gap,0); 
  \draw[arrow] (\x,\h) -- (\x+0.5*\gap,\h+0.5*\h);
  \draw[arrow] (\x+0.5*\gap,\h+0.5*\h) -- (\x+\gap,\h); 
}
  \draw[arrow] (5*\gap,\h) -- (5*\gap,0); 
  \draw[arrown] (-0.25*\gap,-0.25*\h) -- (0,0);  
  \draw[arrown] (5*\gap,0) -- (5.25*\gap,-0.25*\h);  \draw[arrown] (-0.25*\gap,\h+0.25*\h) -- (0,\h);  
  \draw[arrown] (5*\gap,\h) -- (5.25*\gap,\h+0.25*\h); 
  \draw[arrowm]
    (5.25*\gap,\h+1.2) -- (-.25*\gap,\h+1.2);
    \node[above] at (2.5*\gap, \h+1.3)
    {$a$};   
  \node at (5.75*\gap,0.6*\h) {$\cdots\cdots$};
  \node at (-0.75*\gap,0.6*\h) {$\cdots\cdots$};
  \node[left] at (2.7*\gap,0.7*\h) {\textcolor{red}{$\alpha$}};
  \fill[red] (2.9*\gap,1.1*\h) circle (3pt);
  \fill[white] (2.4*\gap,-0.4*\h) circle (4pt);
  \draw[red, arrowm] (2.9*\gap,1.1*\h) .. controls (2.8*\gap,1.0*\h) and (2.7*\gap,0.7*\h) .. (2.6*\gap,0.2*\h) 
  .. controls (2.4*\gap, -0.4*\h) and (2.3*\gap, -0.7*\h) .. (2.1*\gap, -0.8*\h);
  \draw[dashed, arrow] (2.1*\gap-0.5*\gap,-0.8*\h) -- (2.1*\gap,-0.8*\h);
  \draw[dashed, arrow] (2.1*\gap,-0.8*\h) -- (2.1*\gap+0.5*\gap,-0.8*\h);
  \fill[red] (2.1*\gap,-0.8*\h) circle (3pt);
\end{tikzpicture}
  \caption{The lattice model in $\mathcal{C}$-SSB phase on a honeycomb stripe. The $\mathcal{C}$ symmetry is implemented by fusing a string of type $a\in \mathcal{C}$ into the strip from its top edge. When $\mathcal{C}$ is the Fibonacci fusion category, each edge carries a two‐dimensional Hilbert space spanned by the orthonormal basis $\{|1\rangle,|\tau\rangle\}$. A local operator $O_i^\alpha$ can be defined as the insertion of an anyon of type $\alpha\in\mL_e$ through the stripe, from the top, reference boundary to a trivial dashed line. }
  \label{fig:fib-chain}
\end{figure}
Following our previous discussion, we first construct a clean lattice model in the $\mathcal{C}$‐SSB phase. The model is essentially constructed by putting the string-net Hamiltonian on a honeycomb stripe. As in the standard string-net model, each edge carries a two‐dimensional Hilbert space spanned by the orthonormal basis $\{|1\rangle,|\tau\rangle\}$, interpreted as the two possible string types on that edge. The Hamiltonian is a sum of commuting projectors: 
\[
H_0 = -\sum_v A_v -\sum_p B_p,
\label{eq:cleanstrip}
\]
where $A_v$ and $B_p$ are the usual vertex and plaquette terms of the string‐net model. Specifically, $A_v$ is the projector onto string configurations at vertex $v$ that satisfy the fusion ("branching") rules:
\begin{align}
A_v
\Big|\,{\begin{tikzpicture}[baseline={([yshift=-.5ex]current bounding box.center)},thick,scale=0.6, every node/.style={scale=1.0},
  arrow/.style={
    line width=1.2pt, 
    postaction={decorate},
    decoration={markings, mark=at position 0.6 with {\arrow{latex}}}
  }
  ]
  \def\n{5}       
  \def\gap{2}     
  \def\h{1.5}     
    \fill[black] (0,0) circle (3pt); 
    \draw[arrow] (0,\h) -- (0,0);  
    \draw[arrow] (0,0) -- (\h,-0.5*\h); 
    \draw[arrow] (-\h,-0.5*\h) -- (0,0); 
    \node[right] at (0+0.1,0.5*\h) {$a$};   
    \node at (-\h/2,-0.65*\h) {$b$};
    \node at (\h/2,-0.65*\h) {$c$};
\end{tikzpicture} }\Big\rangle=
\delta_{ab}^c \Big|\,{\begin{tikzpicture}[baseline={([yshift=-.5ex]current bounding box.center)},thick,scale=0.6, every node/.style={scale=1.0},
  arrow/.style={
    line width=1.2pt, 
    postaction={decorate},
    decoration={markings, mark=at position 0.6 with {\arrow{latex}}}
  }
  ]
  \def\n{5}       
  \def\gap{2}     
  \def\h{1.5}     
    \fill[black] (0,0) circle (3pt); 
    \draw[arrow] (0,\h) -- (0,0);  
    \draw[arrow] (0,0) -- (\h,-0.5*\h); 
    \draw[arrow] (-\h,-0.5*\h) -- (0,0); 
    \node[right] at (0+0.1,0.5*\h) {$a$};   
    \node at (-\h/2,-0.65*\h) {$b$};
    \node at (\h/2,-0.65*\h) {$c$};
\end{tikzpicture} }\Big\rangle.
\label{eq:A_i string-net}
\end{align}
In this work, we assume $N_{ab}^c\in\{0,1\}$ (no fusion multiplicity) to simplify the discussions, so we replace $N_{ab}^c$ by $\delta_{ab}^c$, which equals 1 if the fusion channel exists and 0 otherwise. And
\[
B_p \;=\;\sum_{s\in \C}\frac{d_{s}}{\sum_{t\in \C}d_{t}^{2}}\,B_{p}^{s},
\]
where $B_{p}^{s}$ acts by fusing in a closed loop of $s$‐type strings around plaquette $p$:
\begin{equation}
\begin{aligned}
    &B_p^s \bigg|\,{\begin{tikzpicture}[baseline={([yshift=-.4ex]current bounding box.center)},thick,scale=0.6, every node/.style={scale=1.0},
    arrow/.style={
    line width=1.2pt, 
    postaction={decorate},
    decoration={markings, mark=at position 0.7 with {\arrow{latex}}}
    },
    arrown/.style={
    line width=1.2pt, 
    postaction={decorate},
    decoration={markings, mark=at position 0.85 with {\arrow{latex}}}
    }
  ]
  \def\n{5}       
  \def\gap{2.5}     
  \def\h{1.5}     
    \fill[black] (0,0) circle (3pt); 
    \draw[arrow] (0,\h) -- (0,0);  
    \fill[black] (\gap,0) circle (3pt); 
    \draw[arrow] (\gap,\h) -- (\gap,0);  
    \node[left] at (0-0.1,0.5*\h) {$a$};
    \fill[black] (0,\h) circle (3pt);
    \node[right] at (\gap+0.1,0.5*\h) {$a'$};
    \fill[black] (\gap,\h) circle (3pt);
  \draw[arrow] (0,0) -- (0.5*\gap,-0.5*\h);
  \draw[arrow] (0.5*\gap,-0.5*\h) -- (\gap,0); 
  \node[below] at (0.25*\gap,-0.35*\h) {$b$};
  \draw[arrown] (-0.25*\gap,-0.25*\h) -- (0,0);  
  \draw[arrown] (\gap,0) -- (1.25*\gap,-0.25*\h);  
  \node[below] at (0.75*\gap,-0.35*\h) {$b$};
  \draw[arrow] (0,\h) -- (0.5*\gap,\h+0.5*\h);
  \draw[arrow] (0.5*\gap,\h+0.5*\h) -- (\gap,\h); 
  \node[above] at (0.25*\gap,\h+0.35*\h) {$c$};
  \draw[arrown] (-0.25*\gap,\h+0.25*\h) -- (0,\h);  
  \draw[arrown] (\gap,\h) -- (1.25*\gap,\h+0.25*\h);  
  \node[above] at (0.75*\gap,\h+0.35*\h) {$c$};
\end{tikzpicture}  }\bigg\rangle
=\bigg|{\begin{tikzpicture}[baseline={([yshift=-.4ex]current bounding box.center)},thick,scale=0.6, every node/.style={scale=1.0},
    arrow/.style={
    line width=1.2pt, 
    postaction={decorate},
    decoration={markings, mark=at position 0.7 with {\arrow{latex}}}
    },
    arrown/.style={
    line width=1.2pt, 
    postaction={decorate},
    decoration={markings, mark=at position 0.85 with {\arrow{latex}}}
    }
  ]
  \def\n{5}       
  \def\gap{2.5}     
  \def\h{1.5}     
    \fill[black] (0,0) circle (3pt); 
    \draw[arrow] (0,\h) -- (0,0);  
    \fill[black] (\gap,0) circle (3pt); 
    \draw[arrow] (\gap,\h) -- (\gap,0);  
    \node[left] at (0-0.1,0.5*\h) {$a$};
    \fill[black] (0,\h) circle (3pt);
    \node[right] at (\gap+0.1,0.5*\h) {$a'$};
    \fill[black] (\gap,\h) circle (3pt);
  \draw[arrow] (0,0) -- (0.5*\gap,-0.5*\h);
  \draw[arrow] (0.5*\gap,-0.5*\h) -- (\gap,0); 
  \node[below] at (0.25*\gap,-0.35*\h) {$b$};
  \draw[arrown] (-0.25*\gap,-0.25*\h) -- (0,0);  
  \draw[arrown] (\gap,0) -- (1.25*\gap,-0.25*\h);  
  \node[below] at (0.75*\gap,-0.35*\h) {$b$};
  \draw[arrow] (0,\h) -- (0.5*\gap,\h+0.5*\h);
  \draw[arrow] (0.5*\gap,\h+0.5*\h) -- (\gap,\h); 
  \node[above] at (0.25*\gap,\h+0.35*\h) {$c$};
  \draw[arrown] (-0.25*\gap,\h+0.25*\h) -- (0,\h);  
  \draw[arrown] (\gap,\h) -- (1.25*\gap,\h+0.25*\h);  
  \node[above] at (0.75*\gap,\h+0.35*\h) {$c$};
  \draw[thick,line width=1.2pt] (0.5*\gap, 0.5*\h) circle (0.5*\h);
  \draw[thick, -{Latex}] ([shift={(10:0.5*\h)}] 0.5*\gap,.5*\h) arc[start angle=-1, end angle=1,radius=0.5*\h];
  \node[right] at (0.5*\gap-0.1,.5*\h) {$s$};
\end{tikzpicture}  }\bigg\rangle .
\end{aligned}
\label{eq:plaquetteterm}
\end{equation}
For a model with string types valued in $\mathcal C$, the lattice model exhibits a symmetry associated with each object $a\in\mathcal C$. This symmetry operator $W_a$ is implemented by fusing a string of type $a$ into the strip from its top edge, as illustrated in Fig.~\ref{fig:fib-chain}; in our example, this realizes the Fibonacci symmetry. Showing that this symmetry commutes with the Hamiltonian $H_0$ is nontrivial and requires the pentagon identity in Eq.~\eqref{eq:pentagon}, but follows the standard argument in the anyon‐chain literature \cite{2025anyonchain}.

In the ground state of $H_0$, the $\mathcal C$ symmetry is completely spontaneously broken. To see this, we follow the argument of Ref.~\cite{levin20}. Since $H_0$ is a sum of commuting projectors, its ground space is the common $+1$ eigenspace of all local terms. A key property of the string-net model is that any string configuration in this subspace can be related by sequential $F$-moves to a fixed set of reference configurations, whose number depends only on the global topology of the lattice and thus determines the ground-state degeneracy. On a strip with periodic boundary conditions along its extended direction, repeated use of the branching rules and $F$-moves reduces every configuration to a single $a$ loop  winding around the strip, for all $a\in \mathcal{C}$. 
The Fibonacci symmetry acts by mixing the amplitudes of the two configurations according to the fusion rule $a\ket{b}=\sum_{c}N_{ab}^c\ket{c}$. Hence the ground-state degeneracy is $|\mathcal C|=2$, and each symmetric ground state spontaneously breaks the anomalous Fibonacci symmetry. In the language of topological holography introduced in last section, a string-net strip model with all edges labeled by objects in $\C$ corresponds to condensing the electric Lagrangian algebra $\mathcal{L}_e$ on both the top and bottom boundaries.

We now introduce local symmetry-breaking disorder into each realization, requiring only that the disorder distribution remain $\mathcal C$-symmetric. One possible realization is
\begin{equation}
    H_I = H_0 + \sum_{i} h_i\,O_{i}^\alpha,
\end{equation}
where $O_i^\alpha$, illustrated in Fig.~\ref{fig:fib-chain}, inserts an anyon of type $\alpha\in\mL_e$ through the stripe near site $i$. In the string‐net model, an anyon is specified by the string types from which it is built, together with consistency conditions that ensure a closed anyon line can be topologically deformed in the ground state, as described in the original work~\cite{Levin05a}. We then annihilate the anyon $\alpha$: (a) at the bottom, on an auxiliary dashed line, which is fixed to carry the trivial string type (see Fig.~\ref{fig:fib-chain}), by projecting the intersection onto the subspace consistent with the fusion rules, \ie onto the trivial-string sector of $\alpha$ at the corresponding vertex\footnote{Any anyon $\alpha$ in the electric Lagrangian algebra always contains the trivial string type. \cite{Levin05a,kitaev2012}}; (b) we can then terminate the other end on the top edge. In the Fibonacci chain, $\alpha$ is the unique non-trivial anyon of $\mL_e$, namely $\tau\bar{\tau}$. Two properties of the tunneling operators $O_i^\alpha$ are essential for our discussion:
\begin{itemize}
    \item In the Fibonacci chain, the operator $O_i^{\tau\bar{\tau}}$ transforms as
$
O_i^{\tau\bar{\tau}} \;\mapsto\;
-\phi^{-2}\,O_i^{\tau\bar{\tau}}
$
under the Fibonacci symmetry: as shown in Eq.~(104) of Ref.~\cite{levin20}, the $\tau\bar{\tau}$ line acquires an eigenvalue $-\phi^{-2}$ upon braiding with the string type $\tau$, which generates the Fibonacci symmetry.\footnote{When $\Z[\C]=\C\boxtimes\bar{\C}$, as in the Fibonacci case, the half-braiding between an anyon $\alpha$ and the defect line $m\in \C$ on the top canonical boundary reduces to the usual braiding between $\alpha$ and $(m,1)$ in the bulk.\cite{Huang:2023pyk}
} Consequently, to ensure that the transformed Hamiltonian has the same probability as the original one, we can choose the random couplings $\{h_i\}$ from a uniform distribution on $[-1,1]$.

\item All terms in the disordered Hamiltonian continue to commute. This property is inherited from the string-net construction, where an anyon line $\alpha$ commutes with the vertex and plaquette operators. Distinct local anyon-tunneling operators also commute, since anyons in a Lagrangian algebra $\mL_e$ have trivial mutual braiding.
\end{itemize}
Since each disorder realization $H_I$ remains a sum of commuting terms, so its ground state(s) are still obtained by minimizing all terms.  Notably, because it does not commute with the Fibonacci symmetry, the random disorder lifts the degeneracy of the two ground states of $H_0$. For system size $L$ large, let us define 
\[
\Delta \;:=\; \sum_i h_i,
\]
which obeys a Gaussian distribution by the central limit theorem, with mean zero and standard deviation $\Omega(\sqrt{L})$.  Consequently, the probability that $H_I$ has a unique gapped ground state with spectral gap larger than 1 is
\begin{equation}
    P\bigl(|\Delta|>1\bigr) \simeq 1 - \mathrm{const}\,\sqrt{\frac{1}{L}}
    \quad (L\to\infty).
\end{equation}
The preceding analysis shows that, in the thermodynamic limit, a single disorder realization has probability 1 of yielding a unique gapped ground state, and is therefore SRE. This stands in sharp contrast to the scenario addressed in Proposition~\ref{prop:averageKWanomaly}, confirming that here the average $\C$ symmetry is anomaly-free.

\section{Anomaly via module categories}
\label{sec:modulecat}

\subsection{Review of module categories}
\label{sec:modulecatmath}
Mathematically, a $1$D gapped phase with $\mathcal C$ symmetry is modeled by a module category $\mathcal N$ over the fusion category $\mathcal C$~\cite{inamura2021topological,inamura2022lattice,garre2023classifying,thorngren2024fusion,bhardwaj2024categorical,bhardwaj2025gapped}. For our purpose, a $\C$-module category $\N$ is a set of labels (i.e. simple objects), such that $\C$ can ``act'' on them. Instead of providing a formal mathematical definition, we give two physically different but closely related descriptions.

First of all, the simple objects in a module category can be viewed as mathematical abstractions of short-range-correlated ground states in the corresponding gapped phase. The module action describes how symmetry defects terminate in the gapped phase.
Mathematically, we have a functor $\triangleright: \C\times \N \rightarrow \N$.  Suppose we label the simple objects in the module category by $\alpha, \beta, \dots$. For a simple $a\in \C$, we have
\begin{equation}
    a\triangleright \alpha = \bigoplus_\beta N^{a\alpha}_\beta \beta.
\end{equation}
Here $N^{a\alpha}_\beta$ are non-negative integers.
In other words, if $N^{a\alpha}_\beta>0$, we have a ``junction", or a Hilbert space $V^{a;\alpha}_\beta$ of dimension $N^{a\alpha}_\beta$. Associativity requires them to satisfy
\begin{equation}
    \sum_c N^{ab}_c N^{c\alpha}_\gamma=\sum_\beta N^{b\alpha}_\beta N^{a\beta}_\gamma.
\end{equation}

A $\C$-module category is said to be indecomposable, if it is not a direct sum of two $\C$-module categories. 

Having defined the module actions, we need to specify how they interplay with the fusion structures of the symmetry category. This is encoded in the following M moves:
\begin{equation}
\begin{tikzpicture}[baseline={(current bounding box.center)}, scale=0.25,>=stealth]
\def\shift{0};
  \draw[thick ] (0,12) node[above]{$a$} -- (3,9) node[pos=0.9, right]{};
    \draw[thick ]  (3,9) -- (6,6) node[right]{$i$};
    \draw[thick ]  (3,12) node[above]{$b$}-- (3,9);
    \draw[very thick, dotted ] (6,6) -- (6,3)node[below]{$\gamma$} ;
\node[draw=none,fill=none] at (3.7,7.1) {$c$}; 
    \draw[very thick, dotted](6,12) node[above]{$\alpha$}-- (6,6) ;
	\end{tikzpicture} =\sum_{\beta,j,k}[M^{ab\alpha}_{\gamma}]_{c, i}^{\beta, jk} 
    \begin{tikzpicture}[baseline={(current bounding box.center)}, scale=0.25,>=stealth]
    \def\shift{0}
    \draw[thick , -] (\shift+0,12) node[above]{$a$}-- (\shift+6,6) node[pos=0.95, right]{$k$} ;
    \draw[thick, ,-]  (\shift+3,12) node[above]{$b$}-- (\shift+6,9) node[pos=0.9, right]{$j$};
    \draw[very thick, dotted,-] (\shift+6,6) -- (\shift+6,4)node[below]{$\gamma$};
	\node[draw=none,fill=none] at (\shift+6.9, 7.6) {$\beta$}; 
    \draw[very thick, dotted, -](\shift+6, 12) node[above]{$\alpha$}-- (\shift+6,6);
\end{tikzpicture},
\label{eq:Msymbol}
\end{equation}
and also the inverse move $M^{-1}$. $M$ is a unitary isomorphism between $V^{ab}_c\otimes V^{c;\alpha}_\gamma$ and $V^{a;\beta}_\gamma\otimes V^{b;\alpha}_\beta$.

\begin{figure}
\begin{center}
  \includegraphics[width=.70\textwidth]{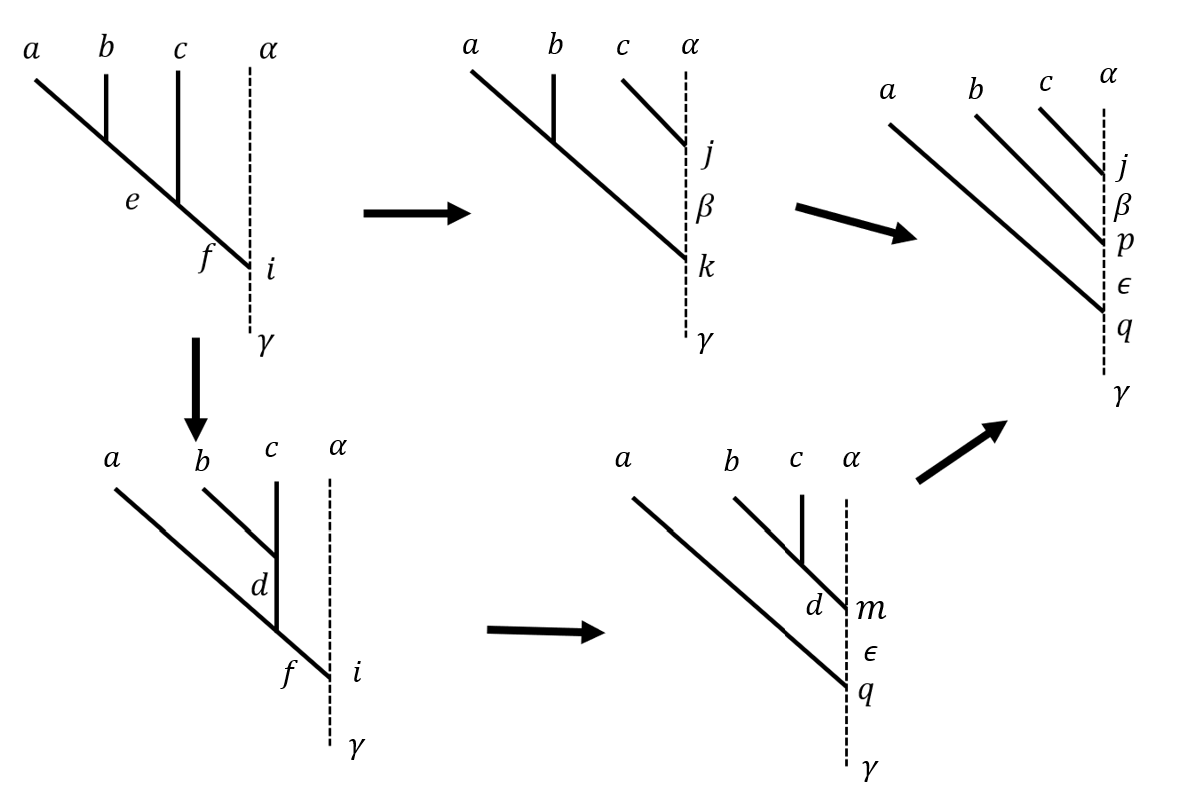} 
\end{center}
\caption{The coupled pentagon equation}
\label{fig:coupledpentagon}
\end{figure}

The M symbols satisfy the coupled pentagon identity:
\begin{equation}
    [M^{ec\alpha}_\gamma]_{f, i}^{\beta,jk} [M^{ab\beta}_\gamma]_{e, k}^{\epsilon,p q} = \sum_{d,m}(F^{abc}_f)_{ed} [M^{ad\alpha}_\gamma]_{f,i}^{\epsilon,mq} [M_\epsilon^{bc\alpha}]_{d,m}^{\beta,jp},
    \label{eq:coupledpentagon}
\end{equation}
which ensures that the two paths in Fig.~\ref{fig:coupledpentagon} yield the same result.

So far we have defined \emph{left} module categories. We can similarly define \emph{right} module categories over $\C$, where $\C$ acts from the right. If a module category $\M$ is equipped with a left $\C$ action, and a right $\D$ action (and they are compatible with each other), then $\M$ is a $(\C, \D)$-bimodule category.

We now give some examples. For any fusion category $\C$, there is a ``canonical" module category, essentially just $\C$ itself: the simple objects are the same, the module action is specified by the fusion rules, and the M symbols are given by the F symbols of $\C$. Physically, this module corresponds to the 1D phase where $\C$ is spontaneously broken to nothing. It is also clear that $\C$ is a $(\C, \C)$-bimodule category.

We are often interested in module categories with a single simple object, corresponding to gapped phases with a unique ground state. Such module categories describe SPT phases protected by $\C$. Mathematically, they are equivalent to tensor functors from $\C$ to ${\rm Vec}$, which are called fiber functors. This can be made explicit as follows: Denote the unique simple object in the module category by $\mathbf{1}$. Then define a functor $F$ as 
\begin{equation}
    F: a\mapsto V^{a;\mathbf{1}}_{\mathbf{1}}.
\end{equation}
One can then verify, using the module $M$ moves and the pentagon identities, that $F$ is indeed a tensor functor. An important property is that the quantum dimension $d_a$ of $a$ must be equal to the dimension of $V^{a;\mathbf{1}}_{\mathbf{1}}$. In particular, $d_a$ must be a positive integer if there exists a fiber functor, which implies that non-integral fusion categories are anomalous \cite{ChangTDL}.

A well-known class of anomaly-free fusion categories is Rep$(G)$ for some finite group $G$. Since it is the category of the isomorphism classes of irreducible linear representations of $G$, one can immediately construct a fiber functor, by mapping each simple object of Rep$(G)$ to the corresponding representation space, and the $M$ symbol are the Clebsch-Gordon coefficients (or 3j symbols). 

A useful way to think about module categories is to consider topological boundaries of Levin-Wen string-net models, as first described in Ref. \cite{kitaev2012}. The string types on the boundary edges are drawn from the module category $\M$ instead of $\C$. The module action and the M moves allow the bulk string types to consistently terminate on the boundary, and one can define boundary plaquette operators using the standard construction. 

In particular, there is a gapped boundary corresponding to $\M=\C$. On this boundary, the  fusion category of equivalence classes of excitations is precisely $\C$. This gives an explicit construction of the electric boundary discussed earlier in Sec.\ref{sec:holo}. In a very similar manner, a $(\C, \D)$-bimodule category can be used to construct an interface between two string-net models with input categories $\C$ and $\D$, respectively \cite{kitaev2012, Lootens:2020mso}.

Below we will also use the notion of relative Degline tensor product between module categories, which will be abbreviated as just tensor product.  If $\M$ is a right $\C$-module and $\N$ is a left $\C$-module, then one can define a linear category $\M\boxtimes_\C\N$~\cite{ENO2009}. For a concrete definition suitable for computations, see Ref. \cite{Barter:2018hjs}. Here we give a physical picture. $\M\boxtimes_\C\N$ can be associated with the 1D gapped phase, obtained by putting the $\C$ string-net model on a strip, with $\M$ as the left boundary and $\N$ as the right boundary. The objects of $\M\boxtimes_\C\N$ are the distinct SRE ground states of the strip model. As an example, if we take $\M=\C$, then the strip is nothing but the topological holography construction and we find $\C\boxtimes_\C\N=\N$, and similarly $\M\boxtimes_\C \C=\M$. Thus $\C$ serves as the ``identity" in tensor product.

If $\M$ is a $(\D, \C)$-bimodule category, and $\N$ is a left $\C$-module category, then $\M\boxtimes_\C \N$ is naturally a left $\D$-module category. Pictorially, we may think of it as fusing a $\D-\C$ domain wall $\M$ to the boundary labeled by $\N$ to get a gapped boundary for the $\D$ string-net model.

With the relative Degline tensor product, all $(\C, \C)$-bimodule categories form a tensor category \footnote{In fact it is a fusion 2-category.}, whose unit is $\C$ itself. Physically, such bimodule categories correspond to topological defects in $\Z[\C]$. We then define invertible bimodule categories as those that have an inverse under the tensor products. They correspond to invertible 0-form symmetries of $\Z[\C]$. It is clear that invertible bimodule categories form a group, called the Brauer-Picard group ${\rm BrPic}(\C)$ of the fusion category $\C$. Our heuristic picture suggests that ${\rm BrPic}(\C)\cong {\rm Aut}\big(\Z[\C]\big)$, which is one of the main theorems proven in \cite{ENO2009}.

Following \cite{ENO2009}, we consider $G$-graded extensions of a fusion category $\D$ from the point of view of module categories. In an extension $\C=\oplus_g \C_g$ with $\D\cong \C_1$, each $\C_g$ is automatically a $(\D,\D)$-bimodule category. In fact, they are all invertible and $\C_g\boxtimes_\D \C_h = \C_{gh}$~\cite{ENO2009}. Thus each extension is uniquely associated with a group homomorphism from $G$ to ${\rm BrPic}(\C)$.

Now suppose $\D$ has a left module category $\N$. Since $\C_g$ is a $(\D, \D)$-bimodule category, $\N_g=\C_g\boxtimes_\D \N$ is also a left $\D$-module category. According to our physical understanding discussed earlier, $\C_g\boxtimes_\D \N$ can be viewed as the result of fusing the $\C_g$ invertible domain wall in $\Z(\D)$ into the gapped boundary described by $\N$. Put it in another way, one applies the $g$ symmetry to the boundary condition. 

In addition, we can also construct ${\rm Ind}_\D^\C(\N)=\C\boxtimes_\D \N$, which is automatically a left $\C$ module category (notice that it is indecomposable as a left $\C$ module, but decomposable as a left $\D$ module). We can see that 
\begin{equation}
    \C\boxtimes_\D \N = \bigoplus_g \C_g\boxtimes_\D \N = \bigoplus_g \N_g.
\end{equation}
Physically, we consider a two-dimensional topological order $\mathcal{Z}[\mathcal{D}]$ enriched by a $0$-form symmetry $G$. The induction ${\mathrm{Ind}}_{\mathcal{D}}^{\mathcal{C}}(\mathcal{N})=\mathcal{C}\boxtimes_{\mathcal{D}}\mathcal{N}$ then gives a gapped boundary of this SET characterized by the $G$-orbit of the gapped boundary $\mathcal{N}$.

\subsection{Lattice model}

To support the second part of our main result, we present an explicit lattice-model construction. Our aims are twofold: (i) to show that, if $\mathcal{A}$ admits a $G$-invariant fiber functor $\mathcal{N}$, one can build a disorder ensemble with full (average) $\mathcal{B}$ symmetry in which every realization is SRE; and (ii) to demonstrate that, when such a fiber functor does not exist, an ensemble exhibiting anomalous features can instead be constructed. By definition, a fiber functor corresponds to a unique gapped ground state that preserves $\mathcal{A}$ symmetry and is therefore in one-to-one correspondence with magnetic Lagrangian algebras in $\Z[\mathcal{A}]$. Our construction thus proves that whenever a $G$-invariant $\mL_m$ in $\Z[\A]$ exists, the symmetry $\mathcal{B}$ is free of average anomaly. For pedagogical clarity, we first analyze the case in which the entire $\mathcal{B}$ symmetry is exact and then reduce the $G$ symmetry to the average level.

\begin{figure}[ht]
  \centering
  \begin{tikzpicture}[thick,scale=1.0, every node/.style={scale=1.0},
  arrow/.style={
    line width=1.2pt, 
    postaction={decorate},
    decoration={markings, mark=at position 0.6 with {\arrow{latex reversed}}}
  },
  arrowm/.style={
    line width=1.2pt, 
    postaction={decorate},
    decoration={markings, mark=at position 0.5 with {\arrow{latex}}}
  },
  modulearrow/.style={
    green!60!black,
    postaction={decorate},
    decoration={markings, mark=at position 0.6 with {\arrow{latex}}}
  },
  dualarrow/.style={
    red,
    line width=1.2pt, 
    postaction={decorate},
    decoration={markings, mark=at position 0.7 with {\arrow{latex}}}
  }
  ]

  \def\n{5}       
  \def\gap{2}     
  \def\h{1.5}     

  \foreach \i in {0,1,2,3,4,5} {
    \pgfmathsetmacro\x{\i*\gap}
    \fill[black] (\x,\h) circle (2pt);  
    \draw[arrow] (\x,0) -- (\x,\h);  
    \node[right] at (\x+0.05,0.5*\h) {$b_{\i}$};
    \fill[green!60!black] (\x,0) circle (2pt);
  }

  \foreach \i in {0,1,2,3,4} {
    \pgfmathsetmacro\xA{\i*\gap}
    \pgfmathsetmacro\xB{\xA + \gap}
    \draw[arrow] (\xA,\h) -- (\xB,\h);                
    \node[above] at ({(\xA+\xB)/2},\h+0.1) {$a_{\i}$};
    \draw[modulearrow] (\xA,0) -- (\xB,0); 
    \node[green!60!black,below] at ({(\xA+\xB)/2},-0.1) {$\alpha_{\i}$};
    \fill[black] ({(\xA+\xB)/2},\h/2) circle (3pt);  
    \node[below] at ({(\xA+\xB)/2},\h/2-0.1) {$g_{\i}$};
    \draw[dualarrow] ({0.7*\xA+0.3*\xB},0) -- ({0.7*\xA+0.3*\xB},-0.5*\h);  
    \fill[red] ({0.7*\xA+0.3*\xB},0) circle (2pt); 
    \node[below] at ({0.7*\xA+0.3*\xB},-0.5*\h) {\textcolor{red}{$\beta_{\i}$}};
  }
  \draw[arrow] (-1,\h) -- (0,\h);  
  \draw[arrow] (5*\gap,\h) -- (5*\gap+1,\h);  
  \draw[modulearrow] (-1,0) -- (0,0);  
  \draw[modulearrow] (5*\gap,0) -- (5*\gap+1,0); 
  \draw[arrowm]
    (5*\gap+1,1.5*\h) -- (-1,1.5*\h);
    \node[above] at (2.5*\gap, 1.55*\h)
    {$s_g$};   
  \node at (6*\gap,0.5*\h) {$\cdots\cdots$};
  \node at (-1*\gap,0.5*\h) {$\cdots\cdots$};
  \draw[->, thick, bend left=30] (2.9*\gap,-0.5*\h) to (2.7*\gap,-0.1);
  \node at (2.9*\gap+0.1,-0.5*\h-0.3) {$Q_b$};
\end{tikzpicture}
  \caption{The symmetry is implemented by fusing a left-oriented $s_{g}\in\mathcal B_{g}$ string in from the top, together with a unitary $U^g$ acts on the $G$ spins at plaquettes. In the clean case, the dual strings in red are fixed to be $\beta_i = \mathbf{1}\in\mathcal{B}^*_{\mathcal{M}}$, thus can be neglected. In the disordered setting, we add $\{ Q_b\}$ terms that project the right half of each module edge onto the subspace with trivial $G$-grading [Eq.~\eqref{eq:disorderedH}]. Consequently, in the ground space, $\beta_i$ carries grading $g_i g_{i-1}^{-1}$.}
  \label{fig:anyonic-chain}
\end{figure}
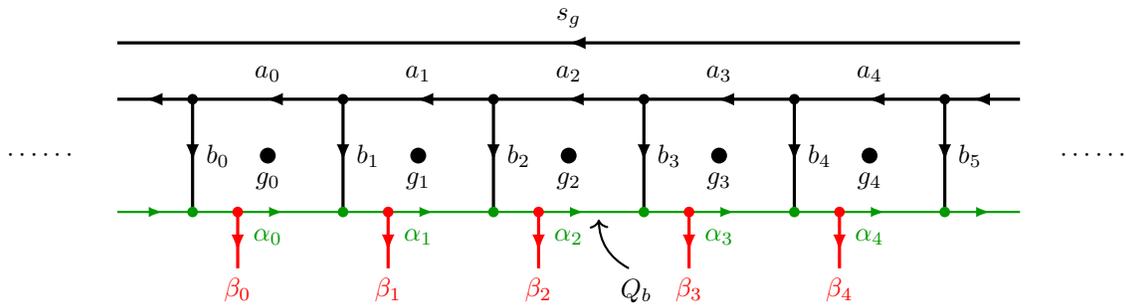

Our construction follows the symmetry-enriched string-net framework introduced in Ref. \cite{ChengPRB2017} and \cite{HeinrichPRB2016}. The model is defined on a square-lattice strip. Local Hilbert spaces are assigned to edges as follows: on each black edge in Fig.\ref{fig:anyonic-chain}, an orthonormal basis (``string type”) is labeled by the objects of $\mathcal B \;=\; \bigoplus_{g\in G} \mathcal B_{g}$. The branching rule of strings obeys the $G$-grading:
\begin{equation}
  \delta_{a_g b_k}^{c_h} = \delta_{gk,h}\,\delta_{a_g b_k}^{c_h}.
  \label{eq:Gstringbranching}
\end{equation} 
By analogy with the previous holographic picture, we will sometimes refer to the black solid lines as the “bulk strings.” By contrast, the green edges at the bottom carry labels drawn from the objects of a $\mathcal B$-module category $\mathcal M$ and will therefore sometimes be referred to as ``module edges'' below. The vector space $V^{a;\alpha_1}_{\alpha_2}$ associated with a trivalent vertex—where a line labeled by $a\in\mathcal B$ meets two lines labeled by $\alpha_1,\alpha_2\in\mathcal M$ and represents the module action—is, in general, multidimensional and is modeled by an internal spin at the vertex. For notational simplicity we omit these vertex spins for the moment—this is justified, \eg when $\mathcal M=\mathcal B$, in which case the vector spaces on the bottom edge are one-dimensional. To incorporate the $G$ symmetry, we place a spin at the center of each plaquette, with basis states $\lvert g\rangle$ labeled by the group elements $g\in G$; these states furnish the regular representation of $G$.

We now introduce the Hamiltonian acting on this Hilbert space: 
\begin{equation}
    H= -\sum_v A_v - \sum_p B_p - \sum_e Q_e,
    \label{eq:Genrichedstringnet}
\end{equation}
where $\{ A_v \}$ are usual string-net vertex projectors, enforcing the string-fusion rules at each vertex on both the top and bottom lines. The edge projectors $\{ Q_e \}$ impose the $G$-grading along each edge \emph{except} the bottom module edge, as illustrated in Fig.~\ref{fig:anyonic-chain}. In particular, since each edge carries an orientation, it has a $G$-spin $g_R$ on its right and $g_L$ on its left. The $Q_e$ term projects the grading of the string to $ g_L g_R^{-1}$. One may regard the region above the top boundary as containing additional $G$-spins, fixed to the state $|1\rangle$ (Here $1$ denotes the identity element of $G$). Finally, $B_p$ denotes the plaquette term, defined as
\begin{equation}
    B_p = \frac{1}{D_\B^2}\sum_{g\in G}|g g_p\rangle\langle g_p|\sum_{s_g\in \B_g} d_{s_g}B_p^{s_g}.
\end{equation}
Here $D_{\mathcal B} = \sqrt{\sum_{s\in\B} d_s^2}$ denotes the total quantum dimension of $\mathcal B$, and $B_{p}^{s_g}$ inserts a closed string of type $s_g$ into plaquette $p$, as defined in Eq.\eqref{eq:plaquetteterm}. For later purposes, we summarize some properties of the plaquette operators here, which are derived in \cite{Levin05a,levin20}
\begin{equation}
    \begin{split}
        & B_p^s B_p^t = \sum_{u\in \B}\delta_{st}^u B_p^u, \\
        & \sum_{s,t\in\B} \delta_{st}^u d_s d_t = D_\B^2 d_u.
    \end{split}
    \label{eq:stringfusion}
\end{equation}
Furthermore, for a $G$-graded fusion category $\B$, we have
\begin{equation}
\label{eq:graded_fusion}
\sum_{\substack{a  \in \B_{g} \\ b\in \B_{h}}} \delta_{ab}^c d_a d_b = D_{\B_1}^2 d_c, \quad \forall c\in \B_{gh},
\end{equation}
as derived in Appendix \ref{app:math}. Here $D_{\B_1} := \sqrt{\sum_{a\in \B_1}d_a^2}=D_\A$ is the quantum dimension of the subcategory $\B_1\cong \A$ with trivial $G$-grading.

The Hamiltonian possesses two important properties: (1) The Hamiltonian is invariant under the global symmetry 
   $\mathcal B=\bigoplus_{g}\mathcal B_g$.  
   The symmetry is implemented by the unitary operator 
   $U^{g}=\bigotimes_{p}U^{g}_{p}$, where each  
   \begin{equation}
       U^{g}_{p}: |g_p\rangle \to| g_p g\rangle
   \end{equation}
acts on the $G$ spin at plaquette $p$, accompanied by fusing a left-oriented $s_{g}\in\mathcal B_{g}$ string in from the top, as depicted in Fig.~\ref{fig:anyonic-chain}. Its commutation with the $A_v$ terms and the $Q_e$ projectors on vertical edges is manifest from their definitions. To show its commutation with the $B_p$ terms, it is sufficient to show that $B_p^s$ commutes with fusing in the $s_g$ line from the top. This is nontrivial and is shown in Ref.\cite{levin20}\footnote{It follows from the proof of the commutativity of the $B_p$ terms in Appendix~C of Ref.~\cite{levin20}.
}, where the pentagon identity of $\B$ is required. Finally, fusing in the left-oriented $s_g$ string while simultaneously flipping the $G$-spins ensures that the symmetry commutes with the $Q_e$ projectors associated with the top horizontal edges. 
 
 (2) The Hamiltonian in Eq.~\eqref{eq:Genrichedstringnet} is a sum of mutually commuting projectors. The operators $A_v$ and $Q_e$ are manifestly projectors. The plaquette operator $B_p$ is also a projector—one can verify the relation $B_p^2 = B_p$ using Eqs.~\eqref{eq:Gstringbranching} and \eqref{eq:stringfusion}. The commutation relations
\begin{equation}
[A_v,Q_e]=0, \quad [A_v, B_p]=0, \quad [Q_e,B_p]=0,
\end{equation}
as well as the commutativity among the $\{A_v\}$ and $\{Q_e\}$ terms, follow immediately from their definitions. The commutativity between the plaquette terms,
\begin{equation}
[B_p,B_{p'}]=0
\end{equation}
is nontrivial and requires the coupled pentagon identity. We will derive this commutation relation in greater detail in the average‐symmetry setting below and in Appendix.~\ref{app:commuting}. We note that when the bottom $\mathcal{B}$-module is chosen to be $\mathcal{B}$ itself, the model reduces to the conventional string-net model (defined on a strip), and the commutativity of all terms follows directly from the standard string-net construction~\cite{Levin05a,levin20}.

Since the Hamiltonian is a sum of commuting projectors, its ground space realizes a gapped phase with $\mathcal{B}$ symmetry, the specific phase depending on the choice of $\mathcal{M}$. We also point out the connection between the lattice model and the discussion in Sec.~\ref{sec:modulecatmath}: the $G$-graded string-net realizes a $G$-enriched topological order, with the underlying anyon theory described by $\Z[\mathcal{A}]$ in the 2D bulk \cite{HeinrichPRB2016,ChengPRB2017}. The ``left'' boundary in the previous section is chosen to be $\mathcal{B}$, and the ``right'' boundary (green line here) is valued in $\mathcal{M}$, thus realizing ground states labeled by objects in $\mathcal{B} \boxtimes_{\mathcal{A}} \mathcal{M}$. The $\mathcal{B}_g$ symmetry defined in this section corresponds to fusing the associated $G$-domain wall in the bulk into the boundary theory realized by the strip model. Moreover, given an $\mathcal{A}$-module category $\mathcal{N}$, fusing bulk strings valued in $\mathcal{B}$ promotes the green module line to take values in the induced $\mathcal{B}$-module category $\M = \mathrm{Ind}_{\mathcal{A}}^{\mathcal{B}}(\mathcal{N})$—thereby capturing all boundary conditions obtained from $\mathcal{N}$ by fusing bulk $G$ domain walls.

We now consider the case in which the full symmetry $\mathcal B$ is an extension of the exact symmetry $\mathcal A$ by a finite group $G$ that is preserved only \emph{on average}. The essential difference in the disordered setting is that the $G$ spins residing on the plaquettes will be treated as quenched disorder: each disorder realization is labeled by their static configuration, collectively denoted by $\{g_p\}$. The disorder ensemble then includes all possible $\{ g_p \}$ configurations.

We first show that, given an $\mathcal{A}$-symmetric SRE phase---i.e., a fiber functor $\mathcal{N}$ of $\mathcal{A}$---which is invariant under $G$, one can construct, for any fixed $G$-spin configuration, a commuting-projector Hamiltonian with a unique gapped ground state in this phase. More precisely, $\mathcal{N}$ being invariant under $G$ means that, after the action of $G$, $\mathcal{N}$ remains equivalent as a $\mathcal{A}$-module category,
\begin{equation}
    \N_g := \B_g\boxtimes_\A \N \simeq_\A \N. 
\end{equation}
Physically, this means that although $\mathcal{N}_g$ may differ from $\mathcal{N}$ in local details after the action of $G$, it remains the same $\mathcal{A}$-SPT phase as $\mathcal{N}$. Combined with Proposition~\ref{prop:averageanomaly}, this establishes our main result: the full symmetry $\mathcal{B}$ is anomaly-free if and only if $\mathcal{A}$ admits a $G$-invariant symmetric SRE phase.

The construction for the disordered setting begins by choosing the green module lines to be drawn from the induced $\mathcal{B}$-module category, i.e.,
\begin{equation}
\mathcal{M} = \mathrm{Ind}_{\mathcal{A}}^{\mathcal{B}}(\mathcal{N}) = \mathcal{B} \boxtimes_{\mathcal{A}} \mathcal{N},
\end{equation}
where $\mathcal{N}$ is the $G$-invariant fiber functor of the exact symmetry $\mathcal{A}$. The trivalent vertices on the module line represent the module action of $\mathcal{B}$ on $\mathcal{M}$, which physically corresponds to a $\mathcal{B}$ domain wall. As a $\mathcal{B}$-module category, these domain walls must fuse consistently with the $F$-symbols of $\mathcal{B}$, thereby satisfying the coupled pentagon identity in Eq.~\eqref{eq:coupledpentagon}. Compared with the clean model, the Hilbert space for the disordered setting includes an additional ancillary string pointing downward from the bottom of each plaquette. The Hilbert space on these ancillary strings takes values in $\mathcal{B}^*_{\mathcal{M}}$, the Morita dual fusion category of $\mathcal{B}$ with respect to $\mathcal{M}$ \cite{2001Ostrik,2002ENO}:
\begin{equation}
    \B_\M^*:=\mathrm{Fun}_\B(\M,\M),
\end{equation}
whose objects are $\mathcal{B}$-module endofunctors from $\mathcal{M}$ to itself, and it is the unique fusion category for which $\mathcal{M}$ is an invertible $\mathcal{B}$–$\mathcal{B}_{\mathcal{M}}^*$ bimodule. For our purposes, the most important property of $\mathcal{B}_{\mathcal{M}}^{*}$ is that, if both $\mathcal{B}$ and $\mathcal{M}$ are $G$-graded, then $\mathcal{B}_{\mathcal{M}}^{*}$ is naturally $G$-graded as well:
\begin{equation}
\begin{split}
    \mathcal{B}_{\mathcal{M}}^{*} \;&=\; \bigoplus_{g\in G} (\mathcal{B}_{\mathcal{M}}^{*})_g,
    \\
    (\mathcal{B}_{\mathcal{M}}^{*})_g
    \;&=\;
    \bigoplus_{h\in G} \mathrm{Fun}_{\mathcal{B}}\bigl(\mathcal{M}_{h},\,\mathcal{M}_{g h}\bigr).
\end{split}   
\end{equation}
As a simple example, if the module category $\mathcal{M}$ is $\mathcal{B}$ itself, then $\mathcal{B}_{\mathcal{M}}^{*} = \mathrm{Fun}_{\mathcal{B}}(\mathcal{B},\mathcal{B}) \cong \mathcal{B}$.

In the disordered setting, the Hamiltonian we consider for a fixed $G$-spin configuration $\{ g_p \}$ is
\begin{equation}
  H(\{g_p\}) \;=-\; \sum_v A_v \;-\; \sum_{e} Q_e \;\; -\; \sum_{p} \widetilde B_p(g_p) -  \;  \sum_b Q_b - \;  \sum_b K_b,
  \label{eq:disorderedH}
\end{equation}
where $\{ A_v \}$, $\{ Q_e \}$ are the vertex and edge terms introduced in Eq.\eqref{eq:Genrichedstringnet}. Each bottom edge $b$ is split into two halves by a vertex that connects to an ancillary edge. The operator $Q_b$ acts on the right half of $b$ and projects onto the subspace in which that segment has trivial $G$-grading, as depicted in Fig.~\ref{fig:anyonic-chain}. Because the $G$-spins are now fixed in each disorder realization and do not evolve dynamically, the plaquette term associated with the plaquette $p$ is modified as follows:
\begin{equation}
  \widetilde B_p(g_p) \;=\;
  |g_p\rangle\langle g_p|\otimes\widetilde B_p,\quad \widetilde B_p:=
  \sum_{s\in\mathcal B_{1}} \frac{d_s}{D_{\mathcal B_1}} B_p^{\,s},
  \label{eq:disordered-plaquette}
\end{equation}
where the sum in $\widetilde B_p$ runs only over string types with trivial $G$ grading. $K_b$ acts on the ancillary string attached to a module edge $b$ and is given by a sum of projectors:
\begin{equation}
    K_b = \sum_{g\in G} K_b^{o_g},
\end{equation}
where $K_b^{o_g}$ projects onto an arbitrary simple object $o_g \in (\mathcal{B}^*_{\mathcal{M}})_g$. For concreteness, we choose $o_1 = \mathbf{1} \in \mathcal{B}^*_{\mathcal{M}}$.

Two properties of the disordered Hamiltonian [Eq.\eqref{eq:disorderedH}] are important for our subsequent discussion:
(1) The ensemble in which all $\{g_p\}$ occur with equal probability realizes the desired symmetry structure, with $\mathcal{A}$ exact and $G$ realized only on average.
It is clear that the $\mathcal{A}$ symmetry—realized by fusing in a string valued in $\mathcal{A}$ from the top without altering the $G$-spin configuration—still commutes with the Hamiltonian for each disorder realization $\{g_p\}$; hence, $\mathcal{A}$ is an exact symmetry of the ensemble. By contrast, an element $\B_g$ with $g\neq \mathbf{1}$ flips each configuration $\{g_p\}$ to $\{g_p g\}$; such operations are therefore respected only at the level of the full disorder ensemble, which includes all possible $\{g_p\}$ configurations with equal probability.

(2) The Hamiltonian in Eq.~\eqref{eq:disorderedH} is a sum of commuting projectors. 
$\{ A_v \}$, $\{ Q_e \}$, $\{ Q_b \}$ and $\{ K_b \}$ are commuting projectors by construction, and their commutation with the $\widetilde{B}_p$ terms is manifest, since the $\widetilde{B}_p$ terms do not change the $G$-grading of the strings. Thus it remains only to verify that the modified plaquette operators $\widetilde B_p$ are themselves commuting projectors.

We begin by showing that each $\widetilde B_p$ is a projector, \ie $\widetilde B_p^2 = \widetilde B_p$. This step parallels the standard string-net analysis and follows directly from the relations in Eq.~\eqref{eq:stringfusion}, where the sum is now taken over the fusion category $\mathcal{B}_1 = \mathcal{A}$. Note that the $\widetilde{B}_p$ terms include the quantum dimension of $\mathcal{A}$ in the denominator. The claim that $[\widetilde B_{p},\widetilde B_{p'}]=0$ is nontrivial and relies on the coupled pentagon equation, Eq.~\eqref{eq:coupledpentagon}, as derived in Appendix \ref{app:commuting}.

The second property above enables an exact solution for any disorder realization $H(\{g_p\})$ in Eq.\eqref{eq:disorderedH}, which we now show possesses a unique gapped ground state. The existence of the gap is immediate: since the Hamiltonian is a sum of commuting projectors, its ground space is a common eigenspace of all terms, and the gap is at least 1. One can also verify that the common $+1$ eigenspace of all projectors is nonempty and therefore forms the ground space. To construct a state in this subspace, choose a single string configuration $|\Psi\rangle$ for which every $\{ A_v \}$, $\{ Q_b \}$, $\{ K_b \}$ and $\{ Q_e \}$ projector takes eigenvalue $+1$.  This can be achieved by choosing the $G$-grading of each string to be aligned with the domain walls in $\{g_p\}$. The role of the ancillary strings valued in $\B_\M^*$ is now evident: they absorb the extra $G$-grading introduced at each domain wall.  In particular, the string on the ancillary edge immediately to the right of a $g$-domain wall carries grading $g$ (see Fig.~\ref{fig:anyonic-chain}), ensuring that the branching rules (i.e., $A_v$ takes eigenvalue $1$) at every vertex are satisfied. A state in which all projectors in Eq.~\eqref{eq:disorderedH} have eigenvalue $1$ is therefore proportional to $\prod_{p}\widetilde{B}_{p}|\Psi\rangle$.

To establish uniqueness of ground state for each disorder realization, we introduce an operator that “removes’’ the disorder:
\begin{equation}
    \begin{split}
        &V(\{ g_p \}) := \prod_p V_p, \\
        & V_p: = | 1\rangle\langle g_p| \sum_{s\in \B_{g_p^{-1}}}  \frac{d_s}{D_\A^2}  B_p^s,
    \end{split}
\end{equation}
where for notational simplicity, we have omitted the fixed grading label $g_p^{-1}$ on the string $s$. Two properties of the operators $\{ V_p \}$ are crucial for our discussion: (1) for a fixed $G$-spin state $|g_p\rangle$, each $V_p$ is invertible on the subspace satisfying $\widetilde{B}_p = 1$:
\begin{equation}
\begin{split}
    (V_p^\dagger V_p) \Pi & =\Bigg( | g_p \rangle \langle g_p| \sum_{\substack{s  \in \B_{g_p} \\ t\in \B_{g_p^{-1}}}} \frac{d_s d_t}{D_\A^4} B_p^s B_p^t\Bigg)\Pi  =( | g_p \rangle \langle g_p| \otimes\widetilde B_p)\Pi =\Pi,
\end{split}
\end{equation}
where $\Pi$ is the projector onto the configuration $\{g_p\}$ and onto the subspace in which all $\widetilde{B}_p = 1$, and we have used Eq.\eqref{eq:graded_fusion}:
\begin{equation}
    \sum_{\substack{s  \in \B_{g_p} \\ t\in \B_{g_p^{-1}}}} \frac{d_s d_t}{D_\A^4} B_p^s B_p^t = \sum_{c\in \B_1}\frac{1}{D_\A^4}\Bigg(\sum_{\substack{s  \in \B_{g_p} \\ t\in \B_{g_p^{-1}}}}d_s d_t \delta_{st}^c\Bigg)B_p^c = \sum_{c\in\B_1}\frac{d_c}{D_\A^2} B_p^c.
\end{equation}
(2) $V_p$ and $V_{p'}$ acting on distinct plaquettes commute. The derivation is the same as that leading to $[\widetilde B_p, \widetilde B_{p'}] = 0$, as shown in Appendix~\ref{app:commuting}. This commutativity is guaranteed by the fact that, as a $\mathcal{B}$-module category, $\mathcal{M}$ satisfies the coupled pentagon identity. These two properties imply that $V(\{ g_p \})$ is a \emph{finite-depth isometry} from the ground space (contained in the $+1$ eigenspace of all plaquette operators) to the full Hilbert space; here, finite depth means that $V(\{ g_p \})$ can be expressed as a product of commuting local isometries.

We now conjugate the disordered Hamiltonian by $V(\{g_p\})$ to make the $G$-spins uniform. For notational simplicity, we denote a particular disorder realization $\{g_p\}$ by the index $I$:
\begin{equation}
    \begin{split}
 H_I & \cong \Pi H_I\Pi\cong \Pi H_I V_I^\dagger V_I\Pi \\ & \cong \Pi  V_I^\dagger
 H_I'V_I\Pi
\end{split}
\end{equation}
Here, the symbol $\cong$ indicates Hamiltonians with the same ground space, and $H_I'$ is obtained by commuting $H_I$ with $V_I^\dagger$:
\begin{equation}
    H_I'= - \sum_v A_v - \sum_b K_b - \sum_e Q_e - \sum_b Q_b(g_p) - \sum_p |1\rangle\langle 1| \otimes \widetilde B_p,
    \label{eq:disorderremoved}
\end{equation}
where we have used that (i) $\{ A_v \}$, $\{ Q_e \}$, $\{ K_b \}$, and $\{ \widetilde{B}_p \}$ commute with $V_I^{\dagger}$, and the final commutation follows from Eq.\eqref{eq:graded_fusion} (It follows more directly from Eq. \eqref{zmul} in Appendix~\ref{app:math}.); (ii) after commuting with $V_I^{\dagger}$, the $\{ Q_b \}$ terms on the bottom edges become $\{g_p\}$-dependent: $Q_b(g_p)$ now projects onto string types in $\mathcal{B}_{g_p^{-1}}$ on the right segment of the bottom edge $b$ of plaquette $p$. One can verify that the Hamiltonian in Eq.~\eqref{eq:disorderremoved} is also a sum of commuting projectors.

Because $V_I$ is an isometry from the subspace defined by $\Pi$—which contains the ground space of $H_I$—to the full Hilbert space, any ground state $|\psi\rangle$ of $H_I$ must map to a ground state of $H_I'$ in which every projector in Eq.~\eqref{eq:disorderremoved} has eigenvalue $1$:
\[
\langle\psi|V_I^{\dagger} H_I' V_I|\psi\rangle
    = \langle\psi| H_I |\psi\rangle .
\]
Consequently, if $H_I'$ has a unique ground state, so does $H_I$.  To establish that the disordered Hamiltonian in Eq.~\eqref{eq:disorderedH} possesses a unique ground state, it therefore suffices to show that $H_I'$ admits exactly one state on which all of its projectors simultaneously take eigenvalue $1$.

We now examine the ground space of $H_I'$. Because all $G$-spins are in the state $|1\rangle$, projecting onto the common $+1$ eigenspace of the $\{ Q_e \}$ terms forces every black solid bulk string to carry trivial $G$-grading, i.e.\ to take values in $\mathcal{B}_1=\mathcal{A}$. Enforcing the branching rules ($\{A_v \}= 1$) together with the $\{ Q_b(g_p) \}$ projectors then fixes the gradings of the module and ancillary strings uniquely for the given disorder realization. Finally, the $\{ \widetilde{B}_p \}$ terms fluctuate the string types; however, the bulk strings remain confined to $\mathcal{A}$. Hence $H_I'$ reduces to a $\mathcal{A}$ string-net model with boundary conditions determined by $\{g_p\}$.

We first consider the case in which $\mathcal{N}$ is a $G$-invariant $\mathcal{A}$ fiber functor; that is, the family $\{\mathcal{N}_g\}_{g \in G}$ consists of equivalent $\mathcal{A}$-fiber functors. For any disorder configuration $\{g_p\}$, the ground space of Eq.~\eqref{eq:disorderremoved} then reduces to a $\mathcal{A}$ string-net strip model whose boundary condition is a $\mathcal{A}$ fiber functor tensored with the $G$ spins in a product state. This model has a unique ground state.\footnote{The $\{ K_b \}$ terms eliminate any residual degeneracy on the ancillary strings by favoring a single string type $o_g$ within each $G$-graded component $(\mathcal{B}_{\mathcal{M}}^{*})_g$.} Consequently, the disordered Hamiltonian in Eq.~\eqref{eq:disorderedH} also possesses a unique ground state. As the unique gapped ground state of a finite-range Hamiltonian, the ground state of Eq.~\eqref{eq:disorderedH} must be $\mathcal{A}$-symmetric and SRE. This completes the proof of the second part of our main result: if $\mathcal{A}$ admits a $G$-invariant fiber functor, then the average categorical symmetry $\B$ is anomaly-free. In particular, one can construct a disorder ensemble with the desired symmetry in which each realization has a symmetric SRE ground state under the exact $\mathcal{A}$ symmetry.

We can further explore the physics that emerges when the average categorical symmetry is anomalous.  As a concrete example, we consider the symmetry  
\[
\mathcal{B} = \mathrm{TY}(\mathbb{Z}_2) = \mathcal{B}_1 \oplus \mathcal{B}_{-1}
\]
in our strip model, with $\mathcal{A} = \mathrm{Vec}_{\mathbb{Z}_2}$ and $G = \mathbb{Z}_2$, which was identified as anomalous symmetry in Sec.~\ref{sec:KWexample}.  The exact symmetry $\mathcal{A}$ admits a fiber functor  
\[
\mathcal{N} \cong \mathcal{B}_{-1} = \{\rho\},
\]
where $\rho$ generates the noninvertible KW transformation and, as an $\mathcal{A}$-module category, corresponds physically to the paramagnet phase in which the exact $\mathbb{Z}_2$ symmetry is unbroken. The module strings then take values in the induced module category  
\[
\mathcal{M} = \mathcal{N}_1 \oplus \mathcal{N}_{-1},
\]
with $\mathcal{N}_1 = \mathcal{N}$ and $\mathcal{N}_{-1} = \mathrm{Vec}_{\mathbb{Z}_2}$, which contains two simple objects representing the two ferromagnetic ground states (all spins up or all spins down).

As discussed above, in the ground space the $\{Q_b(g_p)\}$ terms in $H_I'$ [Eq.\eqref{eq:disorderremoved}] completely fix the $\mathbb{Z}_2$ grading of the module strings according to the realization $\{g_p\}$.  In a generic disorder configuration, the gradings of the module edges become spatially inhomogeneous, alternating between the paramagnetic and $\mathbb{Z}_2$-symmetry-broken ferromagnetic sectors at each domain wall in $\{g_p\}$. 

There are two notable features of this fixed-point ensemble: 
\begin{enumerate}
    \item  A typical disorder realization exhibits extensive ground-state degeneracy. This follows from the fact that each ferromagnetic sector, where the module line takes values in $\N_{-1}$, contributes a factor of 2 to the ground-state degeneracy. Counting the number of ferromagnetic sectors reduces to the classical runs problem for Bernoulli trials \cite{schilling1990longest,feller1991introduction} with probability $p=1/2$ (since the $\bZ_2$-averaged symmetry sets the probability of each spin being up or down to $1/2$), for which the number of ferromagnetic sectors $R$ is asymptotically normal, 
\begin{equation}
    \mathrm{Pr}(R=r) \simeq \sqrt{\frac{8}{\pi(L+1)}}\exp\Bigg[-\frac{8}{L+1}\Big(r-\frac{L+1}{4}\Big)^2\Bigg].
\end{equation}
Consequently, in a typical disorder realization, the number of ferromagnetic sectors scales linearly with system size $L$.

\item The long-range entanglement structure of the ground‐state ensemble can also be inferred.  The circuit depth required to prepare a specific disorder instance $\{g_p\}$ is set by the length of the longest $\mathbb{Z}_2$‐ferromagnetic segment in the system.  The probability that a realization can be prepared by a circuit of depth at most a fixed constant $l$ is therefore the probability that no ferromagnetic segment exceeds length $l$ in the classical runs problem for Bernoulli trials:
\begin{equation}
    \mathrm{Pr}\bigl(\text{depth}\le l\bigr)\;\simeq\;e^{-L/2^{\,l}},
\end{equation}
which is exponentially small in $L$ for fixed $l$.  Here we again use that the average $\mathbb{Z}_2$ symmetry sets the probability of each spin being up or down to $1/2$.  Moreover, for large $L$ the distribution of the longest ferromagnetic segment, and hence of the required circuit depth, is exponentially concentrated around $l=\log_2 L$; the expected number of segments of length $r$ is
\begin{equation}
\mathbb{E}[N_r] \simeq \frac{L}{2^{r+2}}.
\end{equation} 
\end{enumerate}

In summary, in contrast to the anomaly-free case—where we can construct an ensemble of disordered Hamiltonians, each with a unique gapped ground state—when the average categorical symmetry is anomalous, our construction instead produces a disordered ensemble in which a typical realization has extensive ground-state degeneracy and entanglement up to $\log_2 L$ depth.

\section{Conclusions and discussions}

In this work, we present a simple criterion for when an average non-invertible symmetry is anomalous in a one-dimensional disordered system. Specifically, when the full symmetry is a $G$-graded fusion category with the exact symmetry $\A$ as the identity component, the symmetry is anomalous if and only if $\Z[\A]$ does not admit a magnetic Lagrangian algebra that is invariant under the anyon permutation action of $G$, or, equivalently, if $\A$ does not support a $G$-invariant SPT phase. In addition, an average symmetry without any exact component is always anomaly free. We also explore the physical consequences of such average anomalies: in their presence, a typical disorder realization is long-range entangled with probability one in the thermodynamic limit and is expected, on physical grounds, to exhibit power-law Griffiths singularities at low energies. A key conceptual ingredient of our approach is a topological holographic framework that maps one-dimensional systems with (average) categorical symmetries to boundaries of two-dimensional (respectively, average) symmetry-enriched topological phases. Within this framework, anomalies are reinterpreted as obstructions to realizing symmetric gapped boundaries, rendering the classification problem both transparent and computationally tractable. We further complement this perspective with explicit lattice constructions based on symmetry-enriched string-net models, which realize both anomaly-free and anomalous scenarios in a controlled and exactly solvable manner.

We conclude with several open directions.

\emph{Mixed states.} In this work we focus on ensembles of random Hamiltonians with quenched disorder. A natural question is whether exact (strong) and average (weak) non-invertible symmetries can lead to topological phenomena in mixed states that lack a preferred decomposition basis, generalizing recent developments on mixed-state phases with group-like symmetries \cite{2022openSPT,2022ASPT,2025ASPT,2025mixedstateanomaly,2025mixedTO1,2025mixedTO2,2025mixedTO3}. A particularly interesting question is whether a strong anomalous non-invertible symmetry leads to long-range multipartite entanglement, and whether a strong–weak mixed anomaly forbids a gapped Markovian state \cite{2025mixedstateanomaly}. Recently, a topological holographic picture for mixed states has been proposed using the ``doubled space'' technique \cite{2025Qi,2025Schafer,2025Luo}. However, the implications of the mathematical results in doubled space for the physics of the original mixed state remain unclear. Moreover, non-invertible symmetries can also emerge in classical statistical–mechanical models at criticality. It is natural to ask whether average non-invertible symmetries can yield new phenomena even in classical systems.

\emph{Refined constraints on entanglement patterns.} In disordered ground states, a rich variety of quantum entanglement structures arise beyond the conventional short-range versus long-range entangled dichotomy for gapped systems with translation symmetry. Their interplay with symmetry remains largely open. For example, one may ask whether certain anomalous (average) symmetries enforce or forbid multipartite or distillable entanglement \cite{2025distillable} at long distance—indeed, even their precise definitions in many-body systems remain unsettled.

\emph{Fermionic systems and higher dimensions.} In this work, we study the effects of (average) non-invertible symmetry in 1D bosonic systems and map the problem to the boundary of a 2D bosonic (average) SET phase. Recently, classifications of fermionic SET phases have been developed \cite{2021fSET,2022fSET}. It is an interesting question to generalize the study of categorical symmetry to disordered fermionic systems and to investigate its connections with (average) fermionic SETs in 2D; some work has been carried out in clean settings \cite{2024fholo}. Moreover, the topological holographic approach for gapped phases with exact symmetries has been extended to higher dimensions \cite{2017Kong}, and it would be interesting to explore whether average non-invertible symmetries in disordered settings and their associated anomalies admit a natural generalization to higher dimensions, potentially via higher-dimensional topological holography.

\begin{acknowledgements}
We thank Michael Levin, Marvin Qi, Xinping Yang and Rongge Xu for helpful discussions. R.M. acknowledges support from grant NSF PHY-2309135 and the Simons Investigator program (C. Xu). Y.L. was supported by the U.S. National Science Foundation under Grant No. NSF DMR-2316598. MC is partially supported by NSF grant DMR-2424315. 
\end{acknowledgements}

\appendix

\section{A mathematical result on $G$-graded fusion categories}
\label{app:math}

Let $\mathcal{B}=\bigoplus_{g\in G}\mathcal{B}_g$ be a $G$-graded fusion category with trivial component $\mathcal{B}_1$.
For a simple object $x$, let $d_x$ denote its quantum (Frobenius--Perron) dimension, and define
\[
D^2:=\sum_{c\in\mathcal{B}_1} d_c^2.
\]

We would like to prove that 
\begin{equation}
    \sum_{a\in\mathcal{B}_g}\sum_{b\in \mathcal{B}_h} d_ad_bN_{ab}^c=D^2 d_c.
    \label{ddeq}
\end{equation}
This follows from the fact that the quantum dimensions as a vector indexed by simple objects is an eigenvector of the fusion coefficient matrix:
\begin{equation}
    \sum_x N_{xb}^c d_x=d_bd_c.
\end{equation}
Now to prove \eqref{ddeq}, first notice that the summation over $a\in {\cal B}_g$ can be replaced with $a\in {\cal B}$, since the grading on $a$ is enforced by $N_{ab}^c$. Thus
\begin{equation}
    \sum_{a\in\mathcal{B}_g}\sum_{b\in \mathcal{B}_h} d_ad_bN_{ab}^c=\sum_{b\in \mathcal{B}_h}d_b\sum_{a\in\mathcal{B}} d_aN_{ab}^c=\sum_{b\in \mathcal{B}_h}d_b^2d_c=D^2d_c.
\end{equation}
Here we use the fact that $\sum_{b\in {\cal B}_h}d_b^2=D^2$ \cite{etingof2015tensor}, which holds as long as the grading is faithful.

A useful equivalent formulation of this result is the following.
Define the formal sum $z_g:=\frac{1}{D^2}\sum_{a\in\mathcal{B}_g} d_a\, a$. Then it is easy to show that \eqref{ddeq} is equivalent to
\begin{equation}
\
    z_gz_h=z_{gh}.
    \label{zmul}
\end{equation}

\section{Proof that the plaquette terms commute}
\label{app:commuting}
\begin{figure}[ht]
  \centering
  \subfloat[]{%
    \includegraphics[width=0.45\columnwidth]{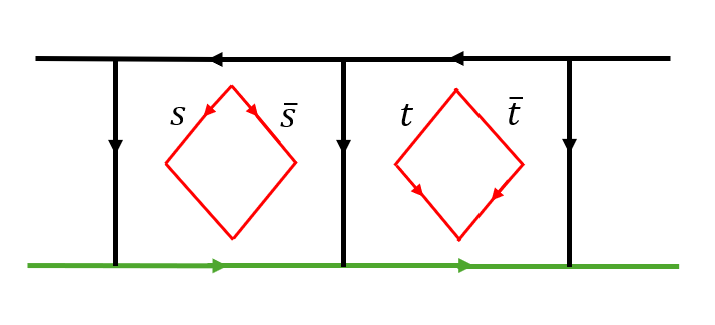}%
    \label{fig:adjacent}%
  }\hfill
  \subfloat[]{%
    \includegraphics[width=0.45\columnwidth]{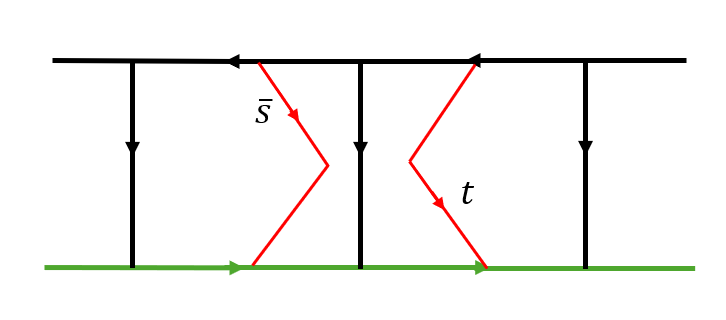}%
    \label{fig:resolve}%
  }
  \caption{The action of $B_p^s$ and $B_{p'}^t$ on the shared edge.}
\end{figure}

Evidently, it suffices to verify $[\tilde{B}_p,\tilde{B}_{p'}]=0$ only when the plaquettes $p$ and $p'$ are adjacent, as depicted in Fig.\ref{fig:adjacent}. After fusing the red loops valued in $\mathcal \B_1$ to the top and bottom lines, the task reduces to checking the commutativity of fusing the $\bar{s}$ and $t$ strings, as shown in Fig.~\ref{fig:resolve}, to the square-lattice edges in two different orders: $\bar{s}$ first or $t$ first, respectively.

\begin{figure}[htb]
  \centering
  \subfloat[]{%
    \includegraphics[width=0.45\columnwidth]{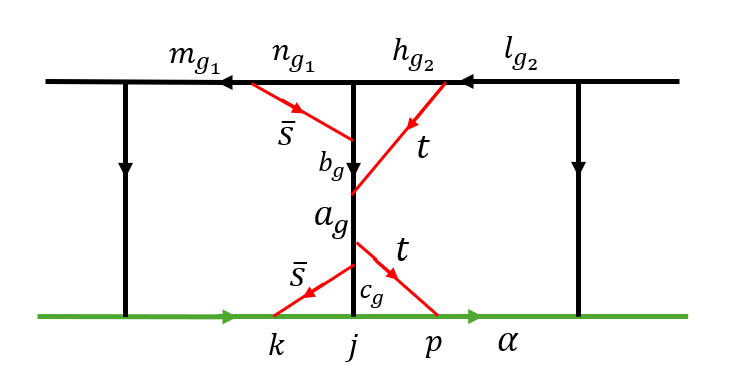}%
    \label{fig:order1}%
  }\hfill
  \subfloat[]{%
    \includegraphics[width=0.45\columnwidth]{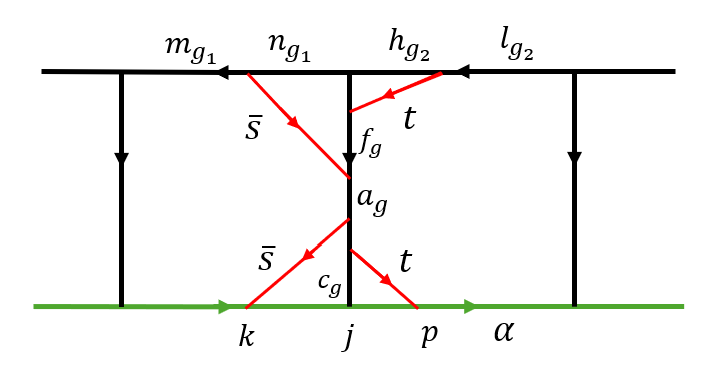}%
    \label{fig:order2}%
  }
  \caption{Two distinct ways of fusing the $\bar s$ and $t$ strings onto the lattice.}
\end{figure}

The first method—i.e., fusing the $\bar{s}$ line in first, as illustrated in Fig.\ref{fig:order1}—produces the following amplitude:
\begin{equation}
  \sum_{b_g,i}
  (\tilde M^{\bar{s}\,c_g\alpha}_\alpha)^{b_g,i}_{\alpha,jk}\,(\tilde M^{b_g\,t\alpha}_\alpha)^{a_g,q}_{\alpha,pi}\,(\tilde F^{m_{g_1} \bar{s}c_{g}}_{h_{g_2}})_{n_{g_1}b_g}(\tilde F^{m_{g_1} b_g t}_{l_{g_2}})_{h_{g_2}a_g},
  \label{eq:leftfirst}
\end{equation}
where: (1) $\tilde{M}$ denotes the inverse of the $M$ move defined in Eq.\eqref{eq:Msymbol}; (2) since the objects in $\M$ are equivalent as $\A$-fiber functors, we denote the string type of the green line by a single label $\alpha$; (3) the $\tilde F$ factor arises from the fusion process at the top trivalent vertex. Here $\tilde F$ denotes the time-reversed $F$-move in Fig.~\ref{fig:fsymbol}), \ie the diagram flipped vertically with all line orientations reversed. Its matrix elements are the complex conjugates of the corresponding $F$-matrix elements, as shown in Ref.~\cite{ChengPRB2017}:
\begin{equation}
    (\tilde F^{abc}_d)_{ef} = (F^{abc}_d)_{ef}^*.
\end{equation}
Conversely, fusing the $t$ line first, as illustrated in Fig.~\ref{fig:order2}, yields the following amplitude:
\begin{equation}
\sum_{f_g,i'}(\tilde M^{c_gt\alpha}_\alpha)_{\alpha,pj}^{f_g,i'} (\tilde M^{\bar{s}f_g\alpha}_\alpha)_{\alpha,i'k}^{a_g,q} (\tilde F^{n_{g_1} c_g t }_{l_{g_2}})_{h_{g_2}f_g}(\tilde F^{m_{g_1}\bar{s} f_g}_{l_{g_2}})_{n_{g_1}a_g},
\label{eq:rightfirst}
\end{equation}
with the $\tilde F$ factor again arising from the fusion process at the top trivalent vertex. It is straightforward to see that, if
\begin{equation}
  \begin{split}
      (\tilde F^{n_{g_1} c_g t }_{l_{g_2}})_{h_{g_2}f_g}(\tilde F^{m_{g_1}\bar{s} f_g}_{l_{g_2}})_{n_{g_1}a_g} & =\sum_{r_g}(\tilde F^{m_{g_1} \bar{s}c_g}_{h_{g_2}})_{n_{g_1}r_g}(\tilde F^{m_{g_1} r_g t}_{l_{g_2}})_{h_{g_2}a_g} (F^{\bar{s}  c_g t }_{a_g})^\dagger_{f_g r_g}, \\
      \sum_{i'}(\tilde M^{c_g t\alpha}_\alpha)_{\alpha,pj}^{f_g,i'} (\tilde M^{\bar{s}f_g\alpha}_\alpha)_{\alpha,i'k}^{a_g,q}&= \sum_{w_g,j'} (\tilde M^{\bar{s}\,c_g \alpha}_\alpha)_{\alpha,jk}^{w_g,j'}\,(\tilde M^{w_g\,t\alpha}_\alpha)_{\alpha,pj'}^{a_g,q}(F^{\bar{s}c_g t}_{a_g})_{w_gf_g}
  \end{split}
  \label{eq:twopentagon}
\end{equation}
holds, substituting these relations into Eq.~\eqref{eq:rightfirst} reproduces Eq.\ref{eq:leftfirst}, thereby demonstrating that the two fusion orders yield identical amplitudes. However, Eq~\eqref{eq:twopentagon} is precisely the pentagon identity of $\mathcal C$, together with the coupled pentagon identity in Eq.~\eqref{eq:coupledpentagon}. We have thus completed the proof that the disordered Hamiltonian [Eq.\eqref{eq:disorderedH}] is a sum of commuting projectors.

\bibliography{average.bib}
\end{document}